\documentclass[fleqn,usenatbib]{mnras}

\usepackage{newtxtext,newtxmath}

\usepackage[T1]{fontenc}
\usepackage{ae,aecompl}

\usepackage{graphicx}	
\usepackage{amsmath}	

\usepackage{amssymb}	

\newcommand{\kms}{km s$^{-1}$}
\newcommand{\Fe}{$^{60}$Fe}
\newcommand{\Pu}[1]{${}^{#1}{\rm Pu}$}

\title[Simulations of \Fe\ from a near-Earth SN]{Simulations of \Fe\ entrained in ejecta from a near-Earth supernova:  Effects of observer motion}

\author[E. Chaikin et al.]{
Evgenii Chaikin,$^{1}$\thanks{E-mail: chaikin@strw.leidenuniv.nl (Evgenii Chaikin)}
Alexander A. Kaurov,$^{2,3,4}$
Brian D. Fields$^{5}$
and Camila A. Correa$^{6}$
\\
$^{1}$Leiden Observatory, Leiden University, P.O. Box 9513, 2300 RA Leiden, the Netherlands\\
$^{2}$Department of the History of Science, Harvard University, Cambridge, MA 02138, USA\\
$^{3}$Blue Marble Space Institute of Science, Seattle, WA 98104, USA\\
$^{4}$Institute for Advanced Study, 1 Einstein Drive, Princeton, NJ 08540, USA\\
$^{5}$Department of Astronomy, University of Illinois, Urbana, IL 61801, USA \\
$^{6}$Institute of Physics, University of Amsterdam, Science Park 904, NL-1098 XH Amsterdam, the Netherlands
}

\date{Accepted XXX. Received YYY; in original form ZZZ}

\pubyear{2021}

\begin{document}
\label{firstpage}
\pagerange{\pageref{firstpage}--\pageref{lastpage}}
\maketitle

\begin{abstract}
Recent studies have shown that live (not decayed) radioactive \Fe{} is present in deep-ocean samples, Antarctic snow, lunar regolith, and cosmic rays. \Fe{} represents supernova (SN) ejecta deposited in the Solar System around $3 \, \rm Myr$ ago, and recently an earlier pulse $\approx 7  \ \rm Myr $ ago has been found. These data point to one or multiple near-Earth SN explosions that presumably participated in the formation of the Local Bubble. We explore this theory using 3D high-resolution smooth-particle hydrodynamical simulations of isolated supernovae with ejecta tracers in a uniform interstellar medium (ISM). The simulation allows us to trace the supernova ejecta in gas form and those eject in dust grains that are entrained with the gas. We consider two cases of diffused ejecta: when the ejecta are well-mixed in the shock and when they are not. In the latter case, we find that these ejecta remain far behind the forward shock, limiting the distance to which entrained ejecta can be delivered to $\approx 100$ pc in an ISM with $n_\mathrm{H}=0.1\; \rm cm^{-3}$ mean hydrogen density. We show that the intensity and the duration of \Fe{} accretion depend on the ISM density and the trajectory of the Solar System. Furthermore, we show the possibility of reproducing the two observed peaks in \Fe{} concentration with this model by assuming two linear trajectories for the Solar System with 30-\kms{} velocity. The fact that we can reproduce the two observed peaks further supports the theory that the \Fe{} signal was originated from near-Earth SNe. 
\end{abstract}




\section{Introduction}

The Milky Way hosts $\sim3$ supernovae (SNe) every century \citep{Tammann1994,Adams2013,Rozwadowska2021}. Such a rate implies the occurrence of at least $1$ SN close to the Solar neighbourhood in the past few Myr.   Indeed, the Sun is surrounded by the Local Bubble, a region 50--150 pc in of low-density, high-temperature gas \citep{Frisch1981,Crutcher1982,Paresce1984,Frisch2011}.  The size and physical conditions of the Local Bubble require multiple supernova explosions over the past $\sim 10 \ \rm Myr$ \citep{Frisch1981,Smith2001,Berghofer2002,Fuchs2006,Abt2011}. Furthermore,  \citet{Breitschwerdt2016} and \citet{Schulreich2017} have argued that the Local Bubble is a superbubble created by $\approx 19$ supernovae over more than 10 Myr timespan.

These astronomical arguments for recent near-Earth supernovae are joined by strong evidence from measurements on and near the Earth and Moon.   The detection of live (not decayed) \Fe, with half-life $2.6$ Myr, demands recent production and delivery to the Earth. \Fe\ has been detected in many deep-ocean ferromanganese crusts and nodules \citep{Knie1999,Knie2004,Fitoussi2008,Wallner2016}, all of which find a pulse
$\approx 3$ Myr ago. The 3-Myr signal also appears in deep-ocean sediments \citep{Fitoussi2008,Wallner2016,Ludwig2016,Wallner2020}, with sufficient time resolution to indicate that the pulse lasted more than $1$ Myr. Very recently \citet{Wallner742} has confirmed a second pulse $\approx 7$ Myr ago. Samples from the Apollo landing show \Fe\ in lunar regolith in excess of cosmic-ray production, and at levels consistent with the deep-ocean measurements \citep{Fimiani2016}. \Fe\ has also been found in cosmic rays \citep{Binns2016}, and in modern Antarctic snow \citep{Koll2019}.

This wealth of \Fe\ data requires an extraterrestrial origin. \Fe{} is largely produced in neutron captures on pre-existing iron in massive stars, mainly in the He and C convective shells  \citep[e.g.][]{2006ApJ...647..483L}. \citet{Fry2015} considered known stellar sources of \Fe\  and concluded that a nearby supernova from a massive star ($\gtrsim 8 \, \rm M_{\rm \odot}$) is the only feasible source. Moreover, the \Fe\ abundance (fluence), together with yields from core-collapse SN models, implies a distance $\approx 20-150$ pc for the SN event 3 Myr ago \citep{Fry2015}. Two star clusters within this distance range have been proposed as candidate sites for this SN: the Tucanae-Horlogium group is closer at $\approx 50$ pc but is relatively small, whereas the Scorpius-Centaurus (Sco-Cen) association is larger and thus more capable of hosting supernovae, but it is more distant at $\approx 100$ pc. \citet{Neuhauser2020} found Sco-Cen to be the likely origin of a runaway pulsar possibly associated with the 3-Myr event.

Several scenarios have been suggested for delivering \Fe\ and other SN ejecta to Earth. \citet{Ellis1996} and subsequent work culminating in \citet{Fry2015} study the `direct deposit' of \Fe\  by a supernova blast engulfing the Solar System, which implicitly is at rest relative to the explosion. \citet{Wallner2016} proposed  that the \Fe\ flux could result from the Sun's motion as it passes through the shell of the Local Bubble. And recently, \citet{Fujimoto2020} studied \Fe\ fluxes resulting from the passage of Milky-Way disk stars through large bubbles of supernova ejecta in spiral arms. As we will discuss in detail below, our work here amounts to a combination of these scenarios.

When \Fe\ arrives in the Solar System, it must be successfully transported to the Earth and Moon in order to be observed, which requires overcoming the repulsion of the solar wind. \citet{Fields2008} showed using gas dynamic simulations that the closest approach of supernova material into the heliosphere is set by ram-pressure balance. Moreover, encroachment of supernova plasma to 1 au requires an explosion at about 10 pc, enough to cause catastrophic biological damage not seen 3 Myr ago, and inconsistent with the distance estimates from the \Fe\ abundance.

But while gas-phase \Fe\ is apparently excluded from $1$ au, iron is a refractory element readily condensed into dust, so that the delivery by dust grains provides a possible mechanism \citep{Benitez2002}. \citet{Athanassiadou2011} and \citet{Fry2016} simulated the transport of SN dust into the (magnetized) solar wind, and found that it is very efficient if grains are large ($> 1 \mu \rm m$) or moving fast, $> v_{\rm esc}(1 \, \rm au) \sim 40$ \kms. \citet{Fry2020} studied charged dust propagation in SN remnants expanding into a magnetized interstellar medium (ISM). They found that the dust particles with sizes of interest quickly decouple from the gas, so at least initially they are not entrained within the gas phase. 

In this study, we use 3D hydrodynamics simulations of an isolated SN in a uniform medium.  We include tracers of \Fe{} that move with the medium and thus  represent {\em entrained} radioisotopes, which either are in the gas phase, or are in the form of dust which come to rest with respect to the medium. We refer to this component of the \Fe{} signal as the {\em entrained flux}. We then simulate the blast properties and the entrained \Fe{} flux for observers at various distances from the explosion. Crucially, we do this for observers moving with a variety of different relative velocities.  We find that a non-zero relative velocity can have important effects on the entrained \Fe{} flux seen by an observer.

Earlier work has examined the dispersal of \Fe{} by supernovae in simulations. \citet{Breitschwerdt2016} and \citet{Schulreich2017} simulated an ensemble of supernovae creating the Local Bubble and dispersion \Fe.
\citet{Fujimoto2020} performed  simulations on Galactic scales. We save a detailed comparison in the discussion section but here we note that our results agree with these important studies in many respects.  Our work is complementary to these studies in that we analyse smaller scales, isolate the effects of a single supernova, consider the effects of diffusion, study different stellar trajectories, and make direct comparison with the latest measured \Fe\ time signatures.

\section{Methods}

\begin{table*}
	\centering

	\begin{tabular}{lllllllll} 
	    Name & $N_{\rm part}$ & $M_{\rm gas}$ [$\rm M_\odot $] & $n_{\rm H}$ [$\rm cm^{-3}$] & $L$ [pc] & $d$ [pc] & $C_{\rm D}$ & Cooling & Initial distribution of \Fe{}  \\
	 
	    \hline
	    \hline
	   \textsf{high\_res\_n01} & $256^3$ & 0.03 & 0.1 & 532 & 2.1 & 0.10 & yes & Uniform, within $r<5.0$ pc\\
	   \textsf{high\_res\_n1} & $256^3$ & 0.09 & 1 & 362 & 1.4 & 0.10 & yes  & Uniform, within $r<3.4$ pc\\
	   \textsf{high\_res\_n001} & $256^3$ & 0.01 & 0.01 & 781 & 3.1 & 0.10 & yes & Uniform, within $r<7.3$ pc\\
	   	\hline
       \textsf{mid\_res\_n01} & $128^3$ & 0.24 & 0.1 & 532 & 4.2 & 0.10 & yes & Uniform, within $r<5.0$ pc\\
       \textsf{low\_res\_n01} & $64^3$ & 1.92 & 0.1 & 532 & 8.3 & 0.10 & yes & Uniform, within $r<5.0$ pc\\
	   \hline
	   \textsf{high\_res\_n01\_highdiff} & $256^3$ & 0.03 & 0.1 & 532 & 2.1 & 0.20 & yes & Uniform, within $r<5.0$ pc\\
	   \textsf{high\_res\_n01\_lowdiff} & $256^3$ & 0.03 & 0.1 & 532 & 2.1 & 0.05 & yes & Uniform, within $r<5.0$  pc\\
	   \textsf{high\_res\_n01\_nodiff} & $256^3$ & 0.03 & 0.1 & 532 & 2.1 & 0.00 & yes & Uniform, within $r<5.0$ pc\\
	   \textsf{high\_res\_n01\_nocooling} & $256^3$ & 0.03 & 0.1 & 532 & 2.1 & 0.10 & no & Uniform, within $r<5.0$ pc\\
	   \textsf{high\_res\_n01\_ej\_rho\_r-2} & $256^3$ & 0.03 & 0.1 & 532 & 2.1 & 0.10 & yes  & $n_{\rm ^{60}Fe} (r) \propto 1/r^2$, within $r<5.0$ pc\\
	   \textsf{high\_res\_n01\_ej\_d2p5pc} & $256^3$ & 0.03 & 0.1 & 532 & 2.1 & 0.10 & yes & Uniform, within $r<2.5$ pc\\
	   \hline

\end{tabular} \\

\caption{Numerical simulations in this work. Column (1) contains names of the simulations; (2) $N_{\rm part}$ is the number of SPH particles in the simulation; (3) $M_{\rm gas}$ is the gas-particle mass; (4) $n_{\rm H}$ is the mean density of the ISM written in units of the number of hydrogen particles per cm$^3$; (5) $L$ is the box size per dimension; (6) $d$ is the average inter-particle distance in the run at the mean ISM density; column (7) indicates the value of the diffusion constant $C_{\rm D}$, which affects the spatial evolution of \Fe{} ejecta. If $C_{\rm D}=0.00$ -- the case of \textit{no diffusion} of \Fe{} ejecta -- then throughout the whole simulation only the particles into which the ejecta were injected as the initial condition carry the ejecta; column (8) shows whether the gas is allowed to radiatively cool; column (9) describes how \Fe{} ejecta are distributed in the initial conditions ($r$ is the distance from the box centre).}
	\label{tab:runs}
\end{table*}

We use 3D numerical simulations of isolated SNe to study the distribution of \Fe{} ejecta in the interstellar medium. We employ the smoothed particle hydrodynamics (SPH) astrophysical code \textsc{swift}\footnote{\textsc{swift} is publicly available at \url{http://www.swiftsim.com}.} \citep{2016arXiv160602738S,2018ascl.soft05020S}, and adopt the energy-density SPH scheme $\textsc{Sphenix}$ \citep{2020arXiv201203974B} to solve the SPH equations of hydrodynamics without including gravitational forces. The $\textsc{Sphenix}$ scheme was originally designed for high-resolution cosmological simulations and has been demonstrated to have excellent performance across various hydrodynamical tests on different scales  \citep{2020arXiv201203974B}. We use the same SPH parameters as in the original paper except for two differences: (1) particle time-steps are limited by the Courant–Friedrichs–Lewy (CFL) parameter $C_{\rm CFL}$ \citep{1928MatAn.100...32C}, which we set to $0.05$ in place of $0.2$, (2) we take the Wendland $C^2$ kernel instead of the Quartic spline. The lower value of the CFL parameter places particles on shorter time-steps leading to the time integration with higher accuracy. The choice of a different SPH kernel has negligible impact on the results presented in this work. Besides the CFL condition, we do not allow the ratio between time-steps of any two neighbouring particles to be greater than $4$. The target smoothing length is set to be
 $1.2348$ times the mean inter-particle separation, which gives the expected number of neighbours $N_{\rm ngb} = 57.28$. 

\subsection{Simulation setup}

We detonate a supernova in a periodic box of homogeneous density\footnote{The distribution of particles in the initial conditions has a glass configuration.} by uniformly distributing $10^{51}$ erg of thermal energy over the SPH particles within a sphere of fixed radius and with the origin coinciding with the box centre. In the simulation at our fiducial resolution (gas-particle mass $M_{\rm gas} = 3 \times 10^{-2} \, \rm M_\odot$) and for our fiducial value of mean ISM density $n_{\rm H} = 0.1$ cm$^{-3}$, this sphere is $5$ pc in radius and includes $57$ particles. 

We then study the evolution of the SN remnant for $4$ Myr. Running simulations for longer times is unnecessary because the first $4$ Myrs suffice to simulate a signal similar to the observed one, while at times greater than $4$ Myr the produced \Fe{} signal generally becomes too weak to be detected. 
To follow the evolution of \Fe, in the initial conditions we uniformly distribute SN-ejecta tracer over the same particles that receive energy from the SN. Following the SN explosion, \Fe{} ejecta propagate outwards through the ISM entrained in the SN blastwave, diffusing and mixing with the swept-up \Fe-free ISM material. 

In order to test how well our results converge, we have also set up runs with the same model but at $8$ and $64$ times lower resolution (by particle mass). In these two runs, the number of particles in the $5$-pc sphere, the region where we distribute SN energy and ejecta, is $9$ and $1$, respectively. These runs are discussed in Appendix \ref{ap:resolution_tests}, in which we study the effects of varying the resolution. Importantly, we show that adjusting the resolution has only a marginal impact on the blastwave properties as well as on the distribution of \Fe{} ejecta in our simulations.

For the ISM gas, we use the temperature floor of $T_{\rm K} = 10^4$ K, corresponding to a typical temperature of the warm ISM, and assume that the gas is fully ionized and has solar metallicity, $Z=0.0134$ and the hydrogen mass fraction $X = 0.737$. The minimum temperature is also the initial temperature of the gas\footnote{The initial temperature of the ISM along with the initial ISM density in our fiducial run and the temperature floor was chosen to resemble the properties of the Local Cloud where the SN is thought to have gone off \citep{Fields2008}. }. Besides varying the resolution, we vary the mean density of the ISM to see how the strength and width of the \Fe{} signal are affected by the environment in which the SN explodes. In our reference run, \textsf{high\_res\_n01}, we take the mean density of the ISM to be equal to $0.1$ hydrogen particles per cubic cm. In addition to this scenario, we consider cases where the mean ISM density $n_{\rm H} = 1.0$ cm$^{-3}$ and $0.01$ cm$^{-3}$. When we increase (decrease) the mean density of the gas by a factor of $10$, we increase (decrease) the gas-particle mass by $0.5$ dex ensuring that the blastwave is always sufficiently far from the edges of the box domain. In these two runs, we keep the same number of particles injected with SN energy and \Fe{} ejecta as in the reference run; to achieve that, in the initial conditions we look for the $57$ particles closest to the box centre. In the initial conditions of the simulation with density $n_{\rm H} = 1.0 \, (0.01)$ cm$^{-3}$, the farthest particle from the box centre containing SN energy and ejecta is at a distance $r=3.41 \, (7.34)$ pc.  We give a full summary of our runs in Table \ref{tab:runs}.

\subsection{Heating and Cooling}

The gas cooling and heating is modelled using precomputed, \textsc{cloudy}-based cooling tables from \cite{2020MNRAS.497.4857P} at redshift $z=0.0$\footnote{We use the fiducial version the cooling tables, \textsf{UVB\_dust1\_CR1 \_G1\_shield} (for the naming convention and more details see Table 5 in \citealt{2020MNRAS.497.4857P}). }. The model of \cite{2020MNRAS.497.4857P} assumes the gas to be in ionization equilibrium in the presence of the ultraviolet/X-ray background of \cite{2020MNRAS.493.1614F}, cosmic rays, and the local interstellar radiation field of a  Milky Way-like galaxy \citep{1987ASSL..134..731B}. The cooling rates of SPH particles are obtained by interpolating the table over density and temperature at a fixed, solar metallicity and redshift $z=0.0$. 

\begin{figure*}
    \centering
    \includegraphics[width=0.8\linewidth]{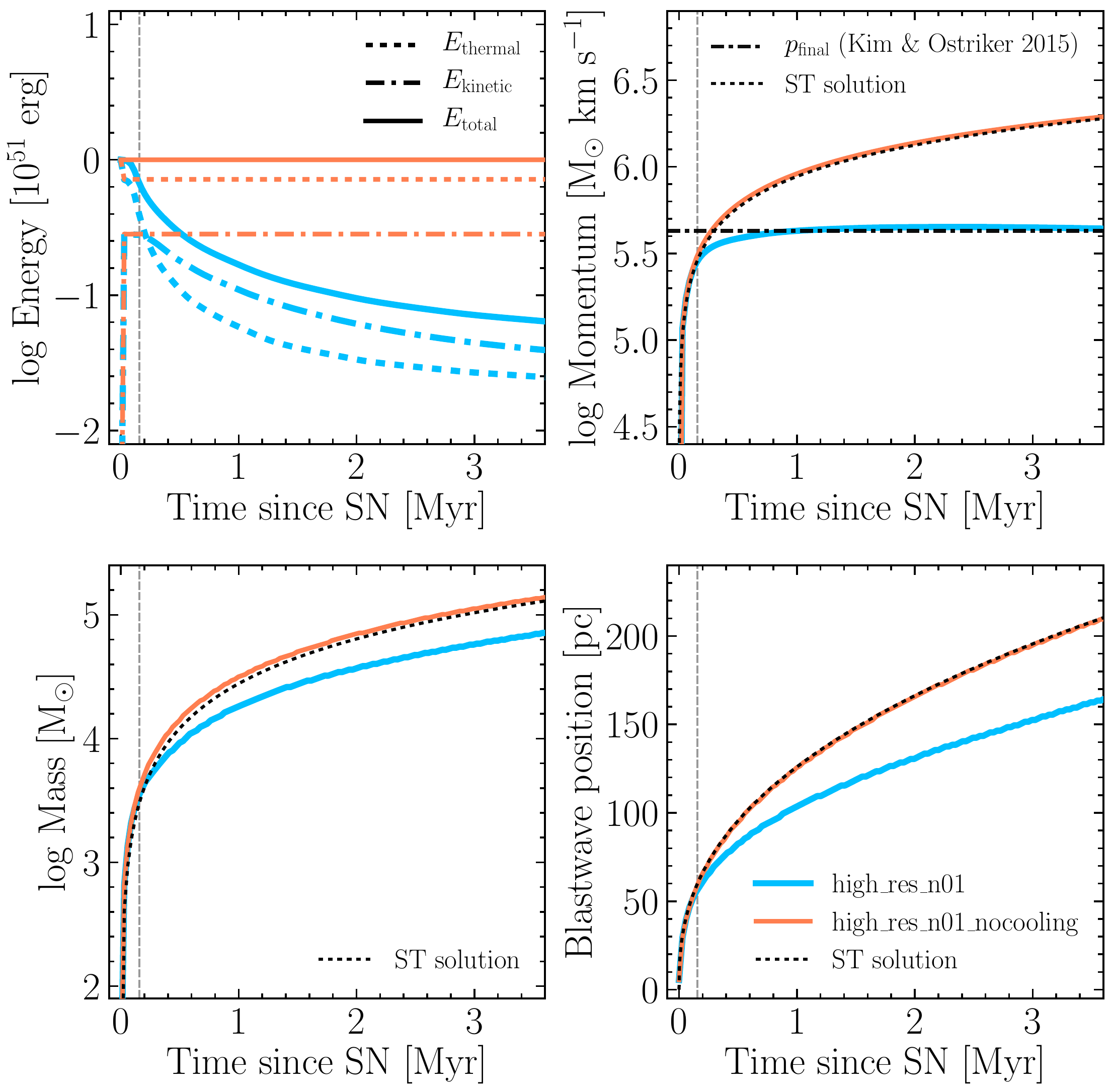}     

    \caption{Properties of the blastwave in the simulations with (\textsf{high\_res\_n01}, blue) and without (\textsf{high\_res\_n01\_nocooling}, orange) gas radiative cooling. The mean ISM density is $n_{\rm H} = 0.1$ cm$^{-3}$ in both cases. \textit{Top left:} Thermal (dashed), kinetic (dash-dotted), and total energy (solid) of the blastwave as functions of time since the detonation. \textit{Top right:} The radial momentum of the blastwave as a function of time. The horizontal, dash-dotted black line is the analytical estimate for the terminal momentum from \citet{2015ApJ...815...67K}. The black, short-dashed curve represents the blastwave radial momentum in the ST phase. \textit{Bottom left:} The total mass inside the blastwave versus time. \textit{Bottom right:} The shock position as a function of time. The vertical, dashed grey line indicates the time when radiation energy losses become important.  }
    \label{fig:energy_and_momentum}
\end{figure*}

Furthermore, we reran the reference model with the gas radiative cooling switched off in order to investigate the impact of cooling on \Fe{}-signal profiles. We also used this run to estimate the accuracy of our simulations and the impact of finite resolution on the signal profiles, both of which were done by comparing the numerical profiles against those derived from the Sedov-Taylor (hereafter ST) self-similar solution to the blastwave problem \citep{1950RSPSA.201..159T,1959sdmm.book.....S}. The differences between the runs with and without radiative cooling are highlighted in Figure \ref{fig:energy_and_momentum} and discussed in $\S$\ref{subsection:numerical_methods}. In the run with the gas cooling switched off, we dropped the minimum allowed temperature from $10^4$ K to $10$ K so that the ISM has negligible thermal pressure, which is one of the assumptions used in the derivation of the ST solution.

\subsection{Analytical solution}
\label{sec:analytical}

In this section we outline the ST analytical solution to the blastwave problem that we use for verification of our numerical methods, as well as for comparison with the earlier works that studied the evolution of \Fe{} during the ST phase \citep{Fry2015,Fry2020}. 

It is well known that for an SN with energy $E_{\rm SN}$, the distance $D$ traversed by the generated blastwave in the energy-conserving phase during time $t$ and in an homogeneous ISM of mass density $\rho$ is given by \citep{1950RSPSA.201..159T, 1988RvMP...60....1O} 

\begin{equation}
    D(t) = \xi_0 \, \left(\frac{E_{\rm SN} t^2}{\rho}\right)^{1/5} \, ,
    \label{eq: distance_SN_time_ST}
\end{equation}
where $\xi_0 = 1.15167$ for the specific-heat ratio of $\gamma = 5/3$. If we require that a blastwave hits the Solar System when $t=2.0$ Myr, we get
\begin{equation}
    D_{\rm SN} = 166 \text{ pc } \left(\frac{E_{\rm SN}}{10^{51} \text{ erg}}\right)^{1/5}\left(\frac{n_{\rm H}}{0.1 \rm \, cm^{-3}} \right)^{-1/5} \left(\frac{t}{2.0 \text{ Myr}}\right)^{2/5} \, ,
    \label{eq:distance_to_SN_sedov_taylor}
\end{equation}
where the hydrogen number density is $n_{\rm H} = \rho \, X / m_{\rm p}$, with $\rm m_{p}$ denoting the proton mass, and $D_{\rm SN}$ should be interpreted as the distance from the Solar System to the SN. The value of $D_{\rm SN}$ should be compared to the so-called fade-away distance \citep{Fry2015}, which is the distance at which the SN shock merges into the ISM, and hence can no longer efficiently bring \Fe{} to the Solar System and Earth,
\begin{equation}
    R_{\rm fade} = 160 \text{ pc }   \left(\frac{E_{\rm SN}}{10^{51} \text{ erg}}\right)^{0.32} \left(\frac{n_{\rm H}}{0.1 \, \mathrm{cm}^{-3}} \right)^{-0.37} \left(\frac{c_{\rm s}}{10 \text{ km s$^{-1}$}}\right)^{-2/5} \, ,
    \label{eq: Fadeaway_distance}
\end{equation}
where $c_{\rm s}$ is the speed of sound. In other words, equation (\ref{eq: Fadeaway_distance}) puts an upper limit on the separation between the Solar System and a hypothetical SN that could deliver \Fe{} onto the Earth surface.

An important aspect of the ST solution is that it may only be applicable until the radiation energy losses in the gas immediately behind the forward shock become important. The latter is given by the time of shell formation \citep[e.g.][]{2015ApJ...815...67K},
\begin{equation}
    t_{\rm sf} = 15.6 \times 10^{4} \text{ yr} \, \left(\frac{E_{\rm SN}}{\text{$10^{51}$ erg}}\right)^{0.22} \left(\frac{n_{\rm H}}{0.1 \, \mathrm{cm}^{-3}}\right)^{-0.55} \, ,
    \label{eq: t_cool}
\end{equation}
which here is written assuming solar metallicity. This shell contains the cool, dense gas that has previously been shocked. At times later than $t_{\rm sf}$, the ST solution (greatly) overestimates the shock radius. We emphasize that the duration of the pulses in the observed \Fe{} data has been consistently reported to be $\gtrsim 1 \, \mathrm{Myr}$, which is much greater than $t_{\rm sf}$. This implies that radiation energy losses cannot generally be neglected when modelling the incorporated rates from the \Fe{} data.

A final consideration is the delivery of SN ejecta to the Earth and Moon, once the Sun encounters supernova ejecta. The Earth and Moon are directly exposed to the SN plasma if this material is carried to 1 au from the Sun by the blast or the motion of the ejecta relative to the Sun. The outgoing solar wind opposes such flows.   \citet{Fields2008}  considered the case of an SN blast sweeping over a stationary Solar System, implicitly with well-mixed ejecta. They showed that we can accurately estimate the closest approach to the Sun by requiring a balance between the ram pressures of the solar wind and the supernova blast. Setting the observed solar wind ram pressure $P_{\rm SW}$ to that for an ST supernova blast, we find that the most distant SN that can envelop the Earth is
\begin{equation}
\label{eq:SN1au}
    R_{\rm 0} = 10 \text{ pc }\left(\frac{E_{\rm SN}}{\text{$10^{51}$ erg}}\right)^{1/3} \left(\frac{P_{\rm SW}}{2 \times 10^{-8}\text{ dyne cm$^{-2}$}}\right)^{-1/3} \, .
\end{equation}
As noted in \citet{Fields2008}, the value in equation (\ref{eq:SN1au}) is close to the $\sim 8 \ \rm pc$ `kill distance' \citep{Gehrels2003,Melott2011} at which supernova can deliver severe damage to Earth's stratospheric ozone layer and thus threaten the biosphere. Thus, supernovae close enough to be dangerous also have blasts strong enough to reach near or within 1 au.

\subsection{Accuracy test. Runs with \& without cooling}
\label{subsection:numerical_methods}

To evaluate the accuracy of our simulations, we compare the temporal evolution of the blastwave produced in the simulations with that from the ST analytical  solution described in \S\ref{sec:analytical}. 

Figure \ref{fig:energy_and_momentum} depicts the energy, momentum, mass, and position of the SN blastwave in our high-resolution runs ($M_{\rm gas} = 3 \times 10^{-2} \, \rm M_\odot$) with and without gas radiative cooling. The mean ISM density in both cases is $n_{\rm H} = 0.1$ cm$^{-3}$. The top left panel displays how the thermal, kinetic and total energies within the blastwave evolve with time. Initially, all the SN energy is injected in thermal form, that is why the kinetic energy is zero. The SN remnant enters an energy-conserving phase where the total energy is conserved, while the thermal energy has decreased by $0.717$ relative to its initial value. In the run without radiative cooling, the energy-conserving phase lasts during the whole time of the simulation, whereas in the model with radiative cooling it ends after $\sim 0.15$ Myr (as predicted by equation \ref{eq: t_cool}). As the gas in the shell of the blastwave cools, the expanding SN remnant loses more of its initial energy to radiation, which slows down the advance of the blastwave and results in the reduced kinetic (and thermal) energy at later times relative to that in the no-cooling scenario.

The top right panel of Figure \ref{fig:energy_and_momentum} displays the temporal evolution of the radial momentum of the blastwave. During the adiabatic, energy-conserving phase, the blastwave momentum rapidly increases. Shortly after the radiation losses start to play a significant role, the blastwave enters the momentum-conserving phase (also known as the `snowplow' phase), where the momentum ceases to grow and approaches an asymptotic value, which in our simulations is in agreement with the value predicted for the terminal momentum by \cite{2015ApJ...815...67K}. Importantly, this panel shows that if the \Fe{} material was deposited onto the Earth surface at a time $\gtrsim 0.2$ Myr since the SN went off and the (average) ISM density was $n_{\rm H} \sim 0.1$ cm$^{-3}$ prior to the explosion, the blastwave approaching the Solar System had to be in the momentum-conserving phase.

The bottom left panel shows how the mass contained within the blastwave increases over time. Compared to the ST solution, our simulation with no cooling losses slightly overpredicts the mass within the shock because the shock front is spread out due to finite resolution.

Finally, in the bottom right panel, we show the position of the forward shock as a function of time. Without radiative cooling, the shock position is exactly the same as that predicted by the ST formula (eq. \ref{eq: distance_SN_time_ST}). To identify the shock position we use radial, equally-spaced bins with a width of $1$ pc. We then approximate the shock position by the radius of the bin that has the maximum radial mass flux ($\rho \, v$). As expected, if the gas is allowed to cool, the shock begins to propagate increasingly slower after the time surpasses the cooling time (eq. \ref{eq: t_cool}), which in the figure is indicated by the vertical dashed lines.

\begin{figure*}
    \centering
    \includegraphics[width=0.97\linewidth]{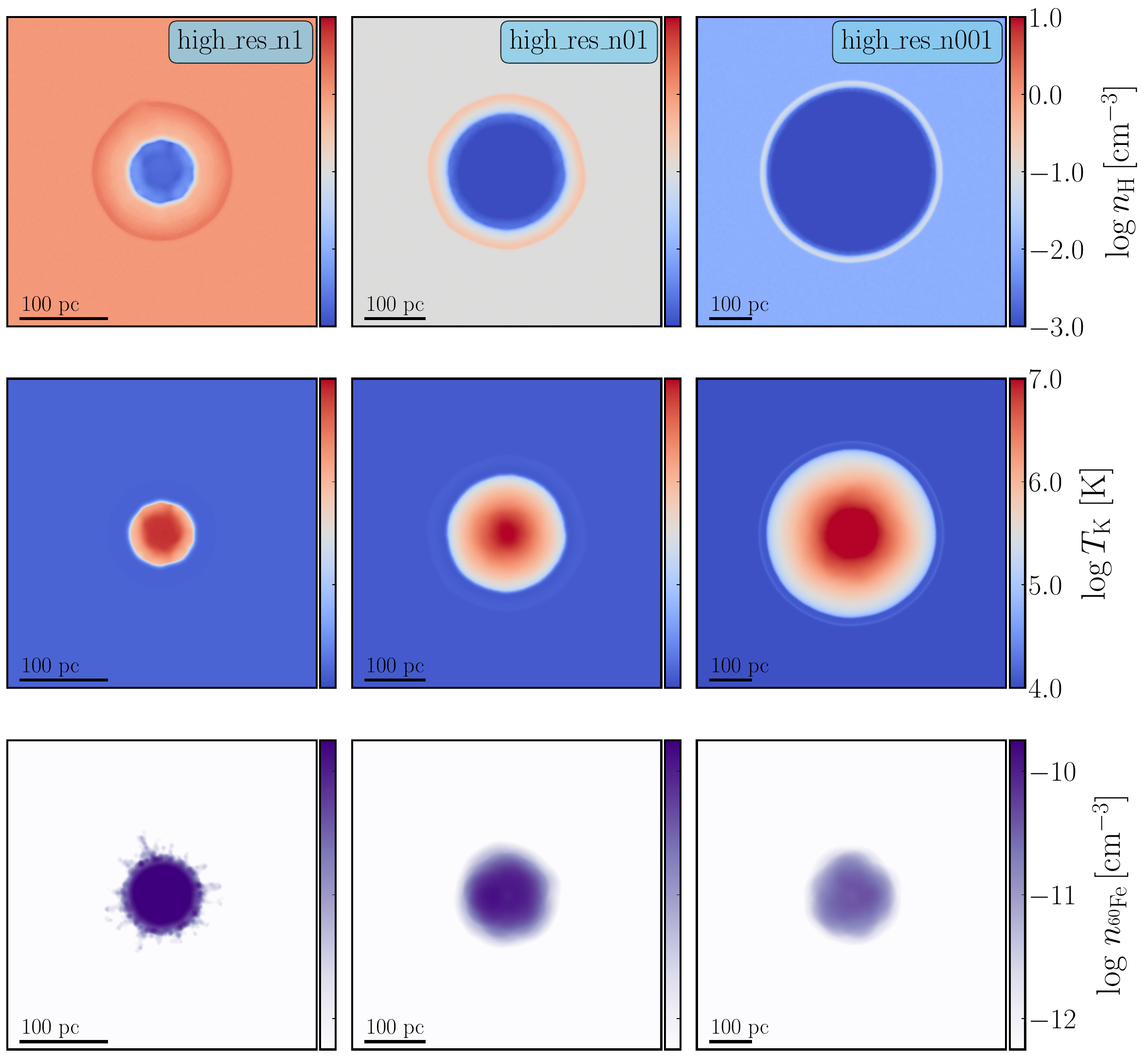}
    \caption{Morphology of the blastwave and \Fe{} ejecta in three simulations with different mean ISM densities: $n_{\rm H} = 1.0$ cm$^{-3}$ (\textsf{high\_res\_n1}, \textit{left column}), $n_{\rm H} = 0.1$ cm$^{-3}$ (\textsf{high\_res\_n01}, \textit{middle column}), and $n_{\rm H} = 0.01$ cm$^{-3}$ (\textsf{high\_res\_n001}, \textit{right column}), shown at $t=2.0$ Myr since the detonation of the SN. Displayed are hydrogen number density (\textit{top row}), gas temperature (\textit{middle row}), and \Fe{} number density (\textit{bottom row}). The fields are averaged in a slab of $20$-pc width whose centre coincides with the position of the SN. To compute the \Fe{} density, we assumed that the amount of \Fe{} produced in the explosion is $10^{-4} \, \rm M_\odot$ and did not account for radioactive decay. The \Fe{} ejecta are only abundant in the hot, low-density bubble far behind the shock front. Each column has its own spatial scale.}
    \label{fig:shock_visualisation}
\end{figure*}

\begin{figure*}
    \centering
    \includegraphics[width=0.99\linewidth]{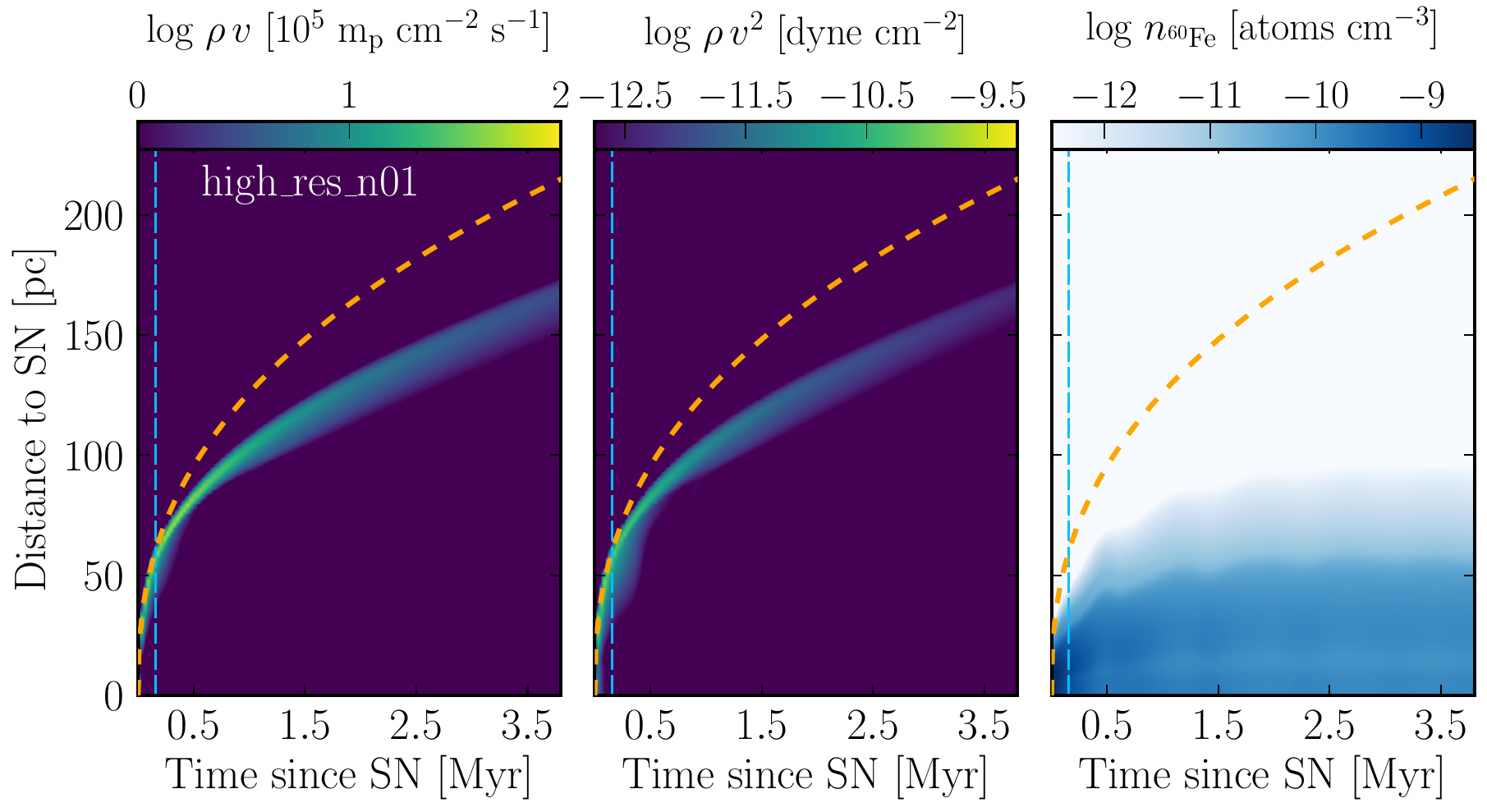}

    \caption{The spherically averaged mass flux (\textit{left}), ram pressure (\textit{middle}) and \Fe-ejecta number density (\textit{right}) in the \textsf{high\_res\_n01} simulation shown as functions of distance to the SN and time since the SN. The vertical, long-dashed light-blue line indicates the moment when the gas cooling losses become relevant. The orange dashed curve stands for the shock position in the ST solution. At a fixed distance, compared to the arrival time of the ST shock, the numerical flux and ram pressure peak later. This is because the advance of the numerical shock is impacted by radiative energy losses. The \Fe{} ejecta lag behind the front of the numerical shock and do not propagate further than $\sim 100$ pc. The right panel assumes a yield $M_{\rm ^{60}Fe, tot}=10^{-4} \, \rm M_\odot$ and no radioactive decay.}
    \label{fig:Fe60_flux_peak_ref_density}
\end{figure*}

\subsection{Ejecta tracing and Diffusion}
\label{sec:ejecta_tracing_and_diffusion}

The ejecta density is traced in all SPH particles. Initially, the ejecta are distributed uniformly among 57 particles surrounding the SN at our fiducial resolution. Then, they propagate to the neighbouring particles according to a diffusion prescription, which is regulated by the diffusion coefficient parameter. 

In our simulations the diffused ejecta are modelled using a subgrid model of turbulent mixing \citep[e.g.][]{2010MNRAS.407.1581S} first proposed by \citet{1963MWRv...91...99S} in modelling of the atmosphere's general circulation. The model has a dimensionless diffusion parameter $C_{\rm D}$ which we set to $0.1$, the value expected from turbulence theory \citep{1963MWRv...91...99S}. Similar values of the diffusion constant have been widely used in the literature \citep{2010MNRAS.407.1581S, 2014MNRAS.443.3809B, 2021MNRAS.506.2836R}. In particular, \citet{2008MNRAS.387..427W} showed that in the standard cluster comparison test \citep[e.g.][]{1999ApJ...525..554F}, using $C_{\rm D}\sim 0.1$ in SPH codes is sufficient to obtain the entropy profiles similar to the predictions by grid codes.

Besides the fiducial model for diffusion ($C_{\rm D}=0.1$), we inspect the \textit{no diffusion} case ($C_{\rm D}=0.0$) as well as two scenarios where the diffusion constant is increased and decreased by a factor of $2$ relative to its fiducial value. Since Lagrangian particles do not exchange mass without a prescription for diffusion, in the case $C_{\rm D}=0.0$ only the particles that received \Fe{} as the initial condition carry it throughout the simulation. These three variations relative to the fiducial model help us estimate the impact of diffusion in the evolution of \Fe{} ejecta (see Appendix \ref{appendix: effect_of_diffusion} for the discussion). Additionally, in Appendix \ref{appendix: effect_of_ICs}, we consider several different initial distributions of \Fe{} and show that variations in the initial conditions have a mild impact on the final distribution of the ejecta.

\begin{figure}
    \centering
    \includegraphics[width=0.99\linewidth]{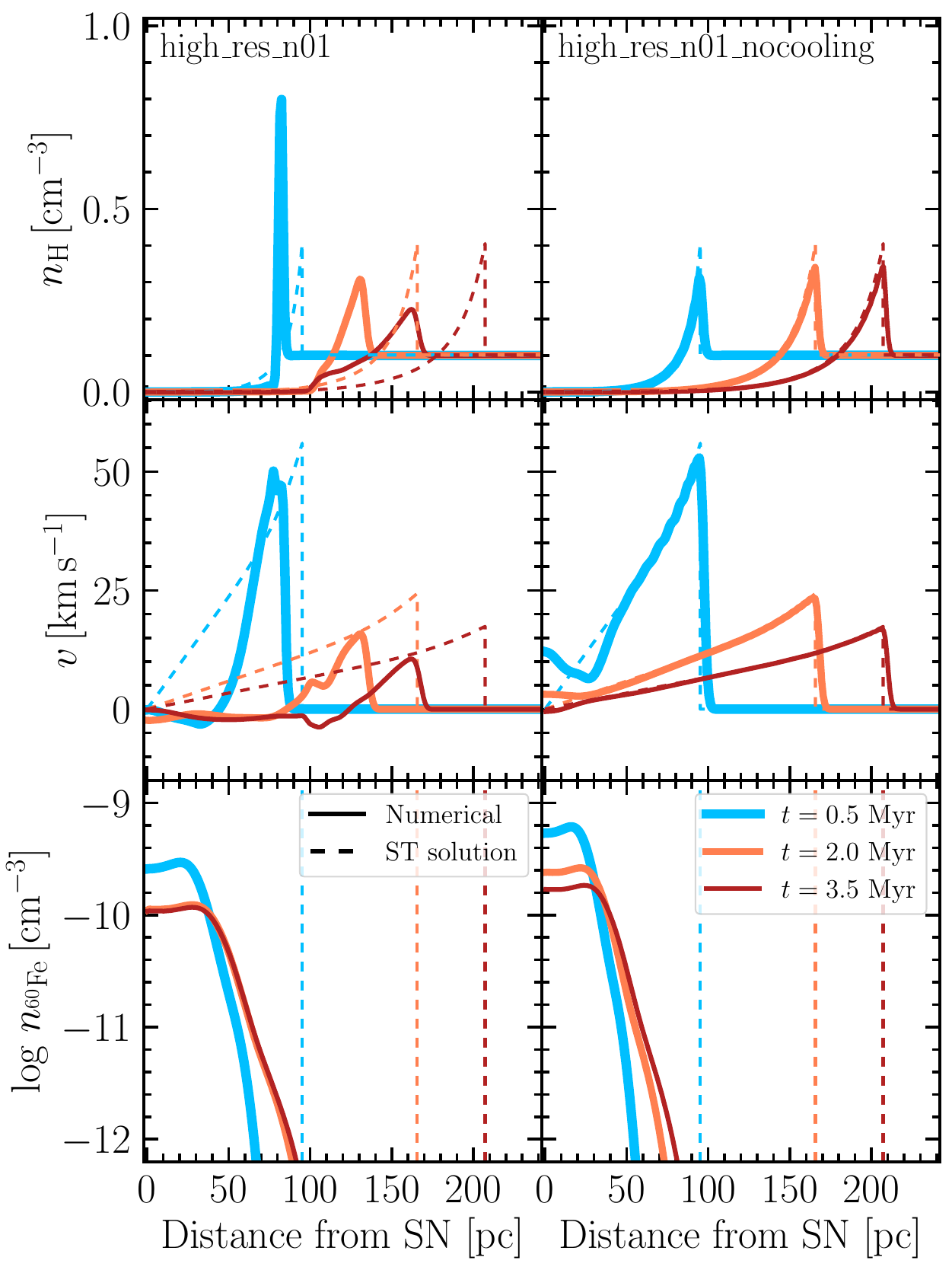}
    \caption{The average radial profiles of gas density (\textit{top}), radial velocity (\textit{middle}), and density of \Fe{} (\textit{bottom}) in the simulations with (\textsf{high\_res\_n01}, \textit{left}) and without (\textsf{high\_res\_n01\_nocooling}, \textit{right}) gas radiative cooling. The profiles are shown at three consecutive times: $0.5$ Myr (light-blue), $2.0$ Myr (orange), and $3.5$ Myr (dark-red). The mean ISM density is $n_{\rm H} = 0.1$ cm$^{-3}$ in both runs. The width of radial bins is $1$ pc.  For reference, the dashed lines show the ST density profiles (\textit{top}), ST velocity profiles (\textit{middle}) and positions of the ST shock (\textit{bottom}). To compute the \Fe-ejecta density we assumed a yield $M_{\rm ^{60}Fe, tot}=10^{-4} \, \rm M_\odot$ and no radioactive decay. The profiles in the run without radiative cooling agree with the ST solution at all times. The ejecta always lag behind the shock front and is always concentrated within a bubble of $100$ pc.}
    \label{fig:hydrogen_and_Fe60_profiles_3times}
\end{figure}

\begin{figure*}
    \centering

    \includegraphics[width=0.95\linewidth]{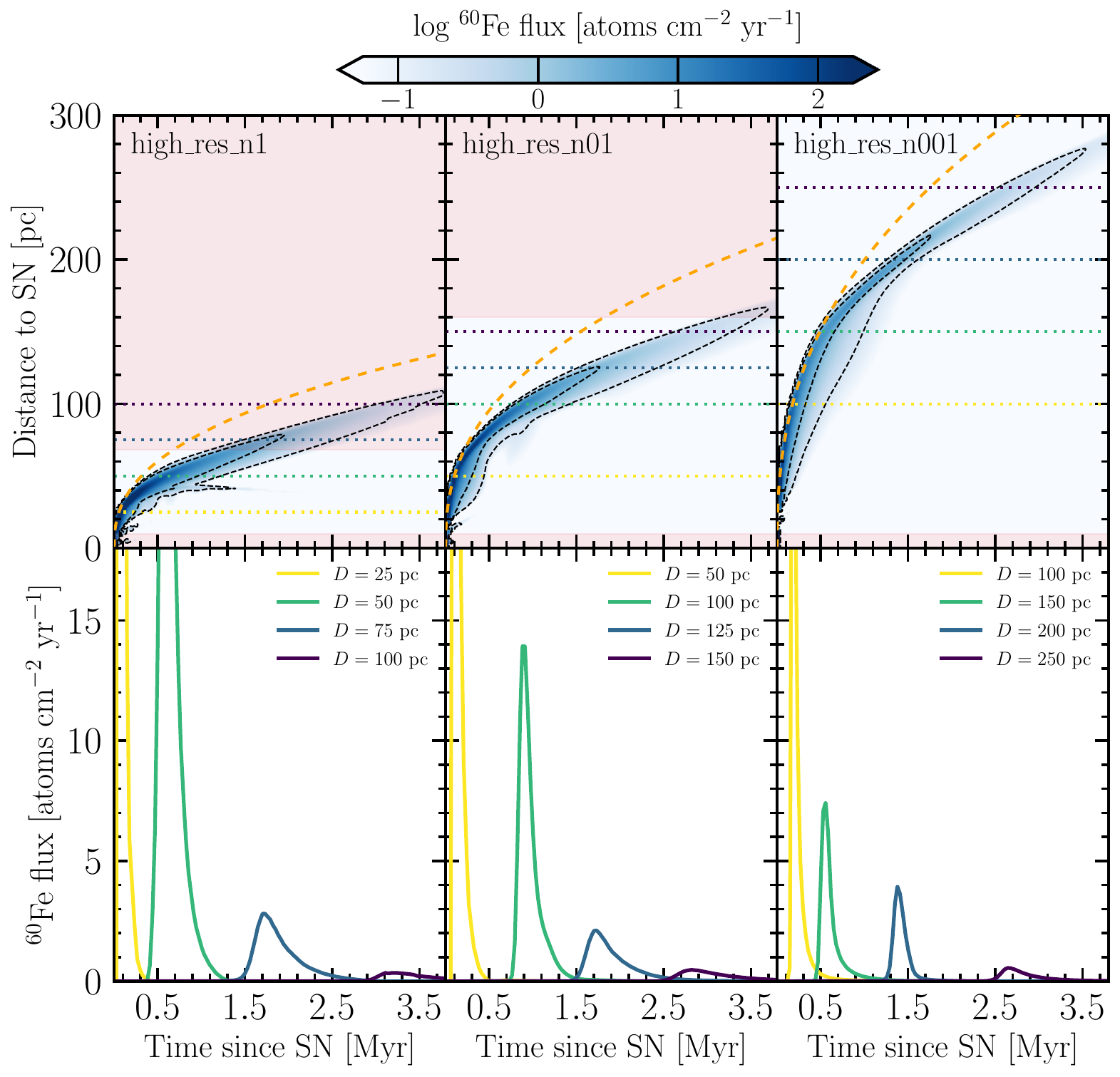}
    \caption{\textit{Top:} Spherically averaged profiles of \Fe{} flux in the well-mixed approximation measured by a static observer shown as a function of distance between the observer and the SN and time of the measurement counted from the detonation of the SN. The profiles are displayed for the high-resolution simulations with mean ISM densities $n_{\rm H}=1.0$ cm$^{-3}$ (\textsf{high\_res\_n1}, \textit{left column}), $n_{\rm H}=0.1$ cm$^{-3}$  (\textsf{high\_res\_n01}, \textit{middle column}), and $n_{\rm H}=0.01$ cm$^{-3}$ (\textsf{high\_res\_n001}, \textit{right column}). The orange dashed curves stand for the shock position in the ST solution. The inner and outer black dashed contours indicate flux values of $2.0$ and $0.2$ \Fe{} atoms cm$^{-2}$ yr$^{-1}$, respectively. The hatched red rectangles indicate the forbidden range of distances between the observer and the SN. They are defined by equations (\ref{eq: Fadeaway_distance}) and (\ref{eq:SN1au}). \textit{Bottom:} The \Fe-flux profiles measured by a static observer located at four different distances from the SN. These distances are shown as horizontal dotted lines in the corresponding top panels. To compute the ejecta flux, we assumed $M_{\rm ^{60}Fe,tot} \, f_{\rm ^{60}Fe} = 10^{-6} \, \rm M_\odot$ and accounted for radioactive decay. }
    \label{fig:flux_well_mix_Fe60}
\end{figure*}

\subsection{Blastwave properties}

\subsubsection{Morphology}

Figure \ref{fig:shock_visualisation} illustrates the morphological characteristics of hydrogen number density, gas temperature, and \Fe-ejecta number density in the SN blastwave in our high-resolution runs at $t=2.0$ Myr for the mean ISM densities of $n_{\rm H} = 1.0$ (left column), $0.1$ (middle column), and $0.01$ cm$^{-3}$ (right column). All fields are averaged in a slab of $20$-pc width centred at the position of the SN. To compute the density of \Fe{}  ejecta, we assume that the amount of \Fe{} produced in the SN is equal to $10^{-4} \, \rm M_\odot$ and do not include corrections due to radioactive decay of the isotope\footnote{In this work, we adopt the following convention: the corrections due to radioactive decay of \Fe{} are never (always) included when computing \Fe{} number density (flux).}. To reconstruct continuous fields from a discrete set of SPH particles and then do the projection of the fields onto a rectangular grid to make images, we used \textsc{Py-SPHViewer}  \citep{alejandro_benitez_llambay_2015_21703}. 

The gas density (top row) and temperature (middle row) reveal that the SN blast has the expected structure: a low-density, hot ($T_{\rm K} \sim 10^{7}$ K) bubble where the gas cooling is unimportant, surrounded by a dense shell where the shocked gas was able to cool down. The gas density monotonically increases from the centre towards the shock front, which defines the outer boundary of the shell. Compared to the gas density, the distribution of \Fe{} ejecta (shown in the bottom row) looks strikingly different. First, the \Fe{} ejecta are noticeably behind the shock front, and they have zero concentration in the shell where the gas density peaks. Second, the ejecta are only abundant inside the hot bubble separated from the much colder and denser shell by a contact discontinuity. Within the bubble, the ejecta density has a roughly flat radial profile with an average value that increases from $n_{\rm ^{60}Fe} \sim 10^{-11} \, \rm cm^{-3}$ at the mean ISM density of $n_{\rm H} = 0.01$ cm$^{-3}$ to more than $10^{-10} \, \rm cm^{-3}$ at $n_{\rm H} = 1.0$ cm$^{-3}$.

The distributions of gas density, temperature and \Fe-ejecta density have similar structures at all three mean ISM densities, except that at the highest mean ISM density, $n_{\rm H} = 1.0$ cm$^{-3}$, we encounter the Rayleigh-Taylor (RT) instability in the ejecta\footnote{The absence of the RT instability in the simulations at mean ISM densities $n_{\rm H} = 0.1$ cm$^{-3}$ and $0.01$ cm$^{-3}$ is due to the (much) smaller gradients in pressure and density across the contact discontinuity.}. The instability develops at the contact discontinuity where the density and pressure gradients are the highest and point to opposite directions, which is a necessary condition for the instability to grow. The overpressured, hot, low-density inner region enriched with \Fe{}  material is strongly pushing against the shell which is dense, cool, and has negligible concentration of \Fe{}. As a result of this interaction, we can see RT `fingers', which effectively increase the size of the \Fe{} bubble by $\sim 10$ per cent.

\subsubsection{Evolution with time}

The left and middle panels of Figure \ref{fig:Fe60_flux_peak_ref_density} show how the spherically averaged mass flux and ram pressure of the SN blast evolve in time since the SN went off. The results are shown only for our fiducial run, \textsf{high\_res\_n01}.  As a reference, the orange dashed line represents the position of the forward shock at a given time in the ST approximation. The ST shock is always ahead of the numerical shock because the latter includes radiative cooling losses. Following the time of shell formation, the width of the flux profile that can be measured by a hypothetical observer at a fixed distance from the SN is within $\sim 1$ Myr. The flux amplitude peaks when the gas in front of the blastwave begins to rapidly cool and become denser. This moment is indicated by the long-dashed, vertical line. The ram pressure and flux peak approximately at the same time and have very similar evolution in both time and space.

For comparison, in the right panel of Figure \ref{fig:Fe60_flux_peak_ref_density} we show how the \Fe{} number density predicted by the simulation changes as a function of time. To compute the \Fe{} density we assumed a yield of $M_{\rm ^{60}Fe, tot}=10^{-4} \, \rm M_\odot$ and ignored radioactive decay (we give the full details of how \Fe{} density and \Fe{} flux are computed in \S\ref{sec:simulationprediction}). We find that the \Fe{} is absent in the shell of the blastwave where the mass flux peaks and instead almost entirely resides behind the shock front, in a low-density bubble with a radius of $\approx 100$ pc. This result is similar to our conclusions from Figure \ref{fig:shock_visualisation}, except that now it is also evident that the \Fe{} bubble reaches its maximum radius at a time $\sim 1$ Myr and then remains roughly constant in time. 

Furthermore, at a given distance smaller than $\approx 100$ pc, the arrival of \Fe{} lags the passage of the blast front because the \Fe{} is not mixed out to the forward shock (at least in our model). This delay between the arrivals of the forward shock and \Fe{} ranges from $\approx 0.0$ Myr for an SN at 20 pc, to $\approx 0.5 $ Myr for an SN at 80 pc.

\subsubsection{Radial profiles}
\label{subsec:oscillations}

The average radial profiles of gas number density, radial velocity and \Fe{} number density are presented in Figure \ref{fig:hydrogen_and_Fe60_profiles_3times}. The profiles are plotted at $0.5$, $2.0$, and $3.5$ Myr after the SN went off. Unlike in Figure \ref{fig:Fe60_flux_peak_ref_density}, here we additionally display the case with no radiative cooling. The figure shows that in the simulation without radiative cooling, the gas density and velocity profiles closely follow analytical predictions. The shock produced in the numerical simulation is slightly spread out relative to the analytical solution but has the correct position at all times, which is an expected behaviour for an SPH code \citep{Rosswog2007, Tasker2008, price2018}. The figure also reveals that in the simulation with radiative cooling, not only the shell expands slower, but also that the radial velocities sometimes become negative. This is the signature of acoustic oscillations that form inside the bubble. This effect can be seen from the oscillating behaviour in the evolution of \Fe{} ejecta in Figure \ref{fig:Fe60_flux_peak_ref_density} and was also reported in similar simulations \citep[e.g.][]{cioffi1988}.

In both cases -- with and without radiative cooling -- we see that the \Fe{} lags behind the shock and is always concentrated within $r < 100$ pc, and the density of \Fe{} becomes negligible in the shock region where the gas density peaks.

\section{Simulation results}

We consider two possible scenarios of the \Fe-ejecta distribution. In \S\ref{sec:wellmixed} we consider the scenario when the ejecta density is proportional to the gas mass density. Then, in \S\ref{sec:simulationprediction} we consider a more realistic scenario when the ejecta are numerically traced as described in \S\ref{sec:ejecta_tracing_and_diffusion}.

\subsection{Well-mixed ejecta}
\label{sec:wellmixed}

In this section we investigate the scenario where the \Fe{} ejecta are \textit{well-mixed} in the SN blastwave; in other words, we assume an instantaneous diffusion within the blastwave that is dominated by the swept-up \Fe-free ISM gas. This is done in order to gain more insight into how the observed \Fe{} signal depends on the ISM properties and in order to ease comparison with the existing literature such as \citet{Fry2015}.

The \Fe{} flux (number of \Fe{} atoms per cm$^{2}$ per yr) reaching the Earth at a time $t$ (counted since the SN went off) can be written as 
\begin{equation}
    F_{\rm ^{60}Fe} =  \frac{1}{4}n_{\rm ^{60}Fe} \, f_{\rm ^{60}Fe}\, |v_{\rm rel}| \, \exp\left(- \frac{t}{\tau_{\rm ^{60}Fe}}\right) \, ,
    \label{eq: flux_Fe_well_mixed}
\end{equation}
where  $n_{\rm ^{60}Fe}$ is the number density of \Fe{} atoms, $|v_{\rm rel}| $ is the absolute relative velocity between the ejecta and the Earth, the exponential factor accounts for the radioactive decay of \Fe{} while the ejecta are on the way to Earth ($\tau_{\rm ^{60}Fe} = 2.6  \, \mathrm{Myr} /  \ln 2$ is the mean lifetime of \Fe), and the dimensionless factor $f_{\rm ^{60}Fe}$ describes the efficiency of the \Fe{} deposition -- the fraction of the released \Fe{} material that is able to penetrate inside the Solar System. In the absence of heliosphere and with a negligible solar-wind pressure,  $f_{\rm ^{60}Fe}$ would be close to one. In reality, however, \Fe{} is expected to be carried to the Solar System by dust grains. In this case, $f_{\rm ^{60}Fe}$ was estimated to be of the order of $0.01$ \citep{Fry2015}. Finally, the factor of $1/4$ is the ratio between the Earth projected surface area ($\pi \rm R_{\earth}^2$) and its total surface area ($4\pi \rm R_{\earth}^2$) with $\rm R_{\earth}$ being the Earth radius. This correction has to be included because even though at a given time the flux is collected by the projected surface area ($\pi \rm R_{\earth}^2$), owing to the Earth rotation \Fe{} should eventually be uniformly spread out over the whole, $4\pi \rm R_{\earth}^2$, surface area.

Under the well-mixed assumption, the number density of \Fe{} atoms is simply proportional to the gas mass density $\rho$,
\begin{equation}
   n_{\rm ^{60}Fe} = \rho \, \frac{M_{\rm ^{60}Fe, tot}}{M_{\rm ej} + M_{\rm ISM}} \frac{1}{\, \mathrm{A_{^{60} Fe}} \, \mathrm{m_u}} \, .
   \label{eq:density_Fe60_well_mixed}
\end{equation}
In the above equation, $M_{\rm ej} \sim 10 \, \rm M_\odot$ is the total mass released by the SN into the ISM (including the part of the metal mass that is $^{60} \rm Fe$), $M_{\rm ^{60}Fe, tot}$ is the total stellar yield of \Fe, $M_{\rm ISM}$ is the mass swept up by the shock by time $t$, $\mathrm{A_{^{60} Fe}}=60$ is the mass number of \Fe, and $\mathrm{m_u}$ is the atomic mass unit. Since we have found that the swept-up mass is $M_{\rm ISM} \gg M_{\rm ej}$ (see the bottom left panel of Figure \ref{fig:energy_and_momentum}), in equation (\ref{eq: flux_Fe_well_mixed}) we safely assume $M_{\rm ej} = 0$.

\begin{figure}
    \centering
    \includegraphics[width=0.98\linewidth]{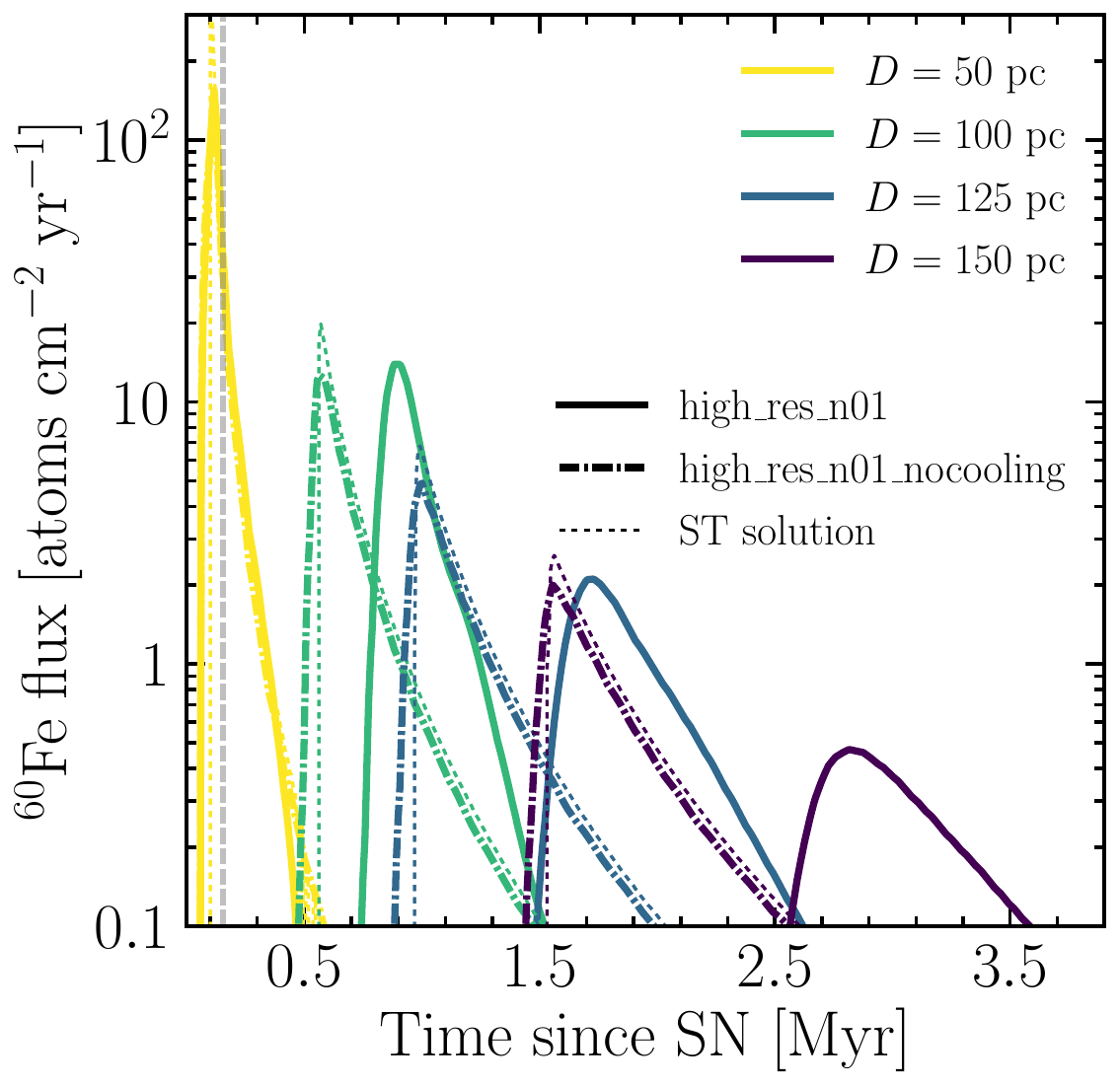}
    \caption{Spherically averaged profiles of \Fe{} flux in the well-mixed approximation measured by a static observer located at several distances from the SN in the simulations with (\textsf{high\_res\_n01}, solid) and without (\textsf{high\_res\_n01\_nocooling}, dash-dotted) gas radiative cooling. The mean ISM density is $n_{\rm H} = 0.1$ cm$^{-3}$ in both cases. For comparison, we additionally show  profiles calculated analytically using the ST solution (thin dashed). To compute the ejecta flux, we assumed that the product $M_{\rm ^{60}Fe,tot} \, f_{\rm ^{60}Fe} = 10^{-6} \, \rm M_\odot$ and accounted for radioactive decay. The vertical, dashed grey line indicates the time when cooling losses become important. }
    \label{fig:flux_well_mixed_with_vs_without_cooling}
\end{figure}

\subsubsection{\Fe{} flux measured by a static observer}

In the well-mixed approximation, the observed \Fe{} flux is described by equation (\ref{eq: flux_Fe_well_mixed}) and the \Fe{} number density is given by equation (\ref{eq:density_Fe60_well_mixed}). If we assume that the observer is \textit{at rest} relative to the position of the SN, then we can study the \Fe{} flux that is spherically averaged (as viewed from the SN position). The latter means that the velocity in equation (\ref{eq: flux_Fe_well_mixed}) is the gas average radial velocity in the reference frame where the SN is at rest. Figure \ref{fig:flux_well_mix_Fe60} displays such radial profiles of  \Fe{} flux in our high-resolution simulations at three mean ISM densities: $n_{\rm H}=1$, $n_{\rm H}=0.1$, and $n_{\rm H}=0.01$ cm$^{-3}$. To normalize the \Fe{} flux, in equation (\ref{eq: flux_Fe_well_mixed}) we took the product $M_{\rm ^{60}Fe,tot} \, f_\mathrm{^{60}Fe}$ to be $10^{-6} \, \rm M_\odot$.  All fluxes are for a static observer.

The panels in the top row illustrate how the flux amplitude and width depend on distance and time. We use the inner and outer black dashed contours to indicate flux values of $2.0$ and $0.2$ \Fe{} atoms per cm$^{2}$ per yr, respectively. For each value of the mean ISM density, we choose four distances to the SN at which the flux is measured by a static observer. For each of these distances, indicated by the vertical dotted lines, we plot the resulting \Fe{} profile in the bottom panel in the same column. For the case of a static observer, it is enlightening to keep track of the constraints on the range of allowed SN distances given by the fade-away argument (eq. \ref{eq: Fadeaway_distance}) and the kill-distance argument (eq. \ref{eq:SN1au}). In short, if the SN goes off at a distance larger than the fade-away distance, then \Fe{} should never arrive on Earth. If the distance to the SN is smaller than the kill distance, then the SN will cause severe damage to the biological life on the planet
\citet{Gehrels2003,Melott2011}. These excluded distances are shown as hatched red regions in the top panels.

Due to the finite resolution of the simulations, the flux onset is not discontinuous, but rather a sharp rise followed by a more gradual falloff.  
We find that the \Fe-flux profiles have typical widths within $1$ Myr and that realistic pulses can easily be constructed using distances from the allowed range of values (i.e. outside the hatched red regions). For example, at mean ISM density of $n_{\rm H} = 0.1$ cm$^{-3}$ and distance of $D = 100$ pc, the \Fe-flux profile has a width of $\sim 0.5$ Myr, which is in agreement with the estimates from \citet{Fry2015}. At a fixed distance, the width of the flux profile is an increasing function of the ISM density. This is because in the higher- (lower-) density medium the blastwave travels slower (faster). The profile shapes are generally asymmetric, with a sharp rise and slower decline, which reflects the density (and velocity) radial profiles that we analysed in Figure \ref{fig:hydrogen_and_Fe60_profiles_3times}. 

\subsubsection{The effect of radiative cooling on signal shape and contribution of smoothing length to signal shape}

Figure \ref{fig:flux_well_mixed_with_vs_without_cooling} compares \Fe-flux profiles in the two high-resolution simulations that were run with and without gas radiative cooling. Both simulations have the mean ISM density of $n_{\rm H} = 0.1$ cm$^{-3}$, and both show results for a static observer. For the latter simulation, it is instructive to contrast the numerical profiles with the ones that are calculated analytically using the ST solution. It can be seen that the profiles from the no radiative cooling simulation are in close agreement with the analytical profiles computed at the same distances. This implies that at our fiducial resolution, the contribution of the SPH smoothing kernel to the overall width of the \Fe{} signal is negligible. 

At a given distance, the inclusion of gas radiative cooling delays the signal, which at late times ($t \gtrsim 1$ Myr) results in a lower amplitude of the pulse. This means that by including radiative energy losses, one can place tighter constraints on the maximum distance to the \Fe{} source. Additionally, gas cooling broadens the signal shape and makes the decline in \Fe{} flux following the peak less rapid.  At the smallest distance between the observer and the SN ($D=50$ pc, in yellow), all three profiles match quite well because the blastwave is still in the energy-conserving phase.

We emphasize that although thus far the cases with and without radiative cooling have both received a considerable amount of attention, the preferred, realistic model is the one including cooling; and this is the only model we will consider in the following.

\subsection{Ejecta as predicted by the simulation}
\label{sec:simulationprediction}
In the previous section, we made the assumption that the shapes of \Fe-ejecta density profile and of blastwave density profile are identical (eq. \ref{eq:density_Fe60_well_mixed}). In general, this is not the case because as the blastwave expands into the ISM, it sweeps up the \Fe-free ISM gas so that the outer layers of the blastwave always contain far smaller concentration of \Fe{} than the hot interior, if at all. In other words, the ejecta always lag behind the front of the forward shock, as we have demonstrated in Figures \ref{fig:shock_visualisation}, \ref{fig:Fe60_flux_peak_ref_density} and \ref{fig:hydrogen_and_Fe60_profiles_3times}.

In order to compute radial profiles of the number density of \Fe{} from the simulations (shown in Figures \ref{fig:Fe60_flux_peak_ref_density} and \ref{fig:hydrogen_and_Fe60_profiles_3times}), we first divided the space into radial bins of the same width centred at the SN location. For a given radial bin $j$, the number density of \Fe{} atoms in bin $j$, $n_{\mathrm{ ^{60}Fe,}j}$, was then calculated as
\begin{equation}
    n_{\mathrm{ ^{60}Fe},j} = \frac{1}{\Delta V_{j}\, \mathrm{A_{^{60} Fe}} \, \mathrm{m_u}}\, \sum_{i} w_{ji} \, M_{\mathrm{gas},i} \, X_{\mathrm{ ^{60}Fe,}i}   \, ,
    \label{eq: n_Fe60_average}
\end{equation}
where the sum is computed over SPH particles whose kernels overlap with the volume of radial bin $j$, $M_{\mathrm{ gas},i}$ is the mass of gas particle $i$, $X_{\mathrm{ ^{60}Fe,}i}$ is particle's mass fraction of \Fe, and $\Delta V_{j}$ is the volume of bin $j$. Finally, the weight $w_{ji}$ is computed as an 3D integral of the Wendland $C^2$ smoothing kernel $W(r)$ over the overlapped region between the extent of the kernel of gas particle $i$ and the volume of bin $j$. Since the kernel function is normalized, if a particle is entirely within a bin, its weight coefficient is equal to one. By construction, the above definition of \Fe{} number density satisfies conservation of mass
\begin{equation}
    \sum_{j} \,  n_{\mathrm{ ^{60}Fe,}j} \Delta V_{j} \, \mathrm{A_{^{60} Fe} \, \mathrm{m_u}} = M_{\rm ^{60}\mathrm{Fe},tot} \, ,
\end{equation}
up to an integration error. Note that in equation (\ref{eq: n_Fe60_average}) we do not need to include the exponential factor due to the decay of \Fe{} because we have already accounted for it in the expression for the flux (eq. \ref{eq: flux_Fe_well_mixed}).

\subsubsection{Role of the Solar System trajectory in the \Fe{} flux}
\label{sec:trajectories}

\begin{figure}
    \centering
    {\large Moving vs. static observer}
    \includegraphics[width=0.97\linewidth]{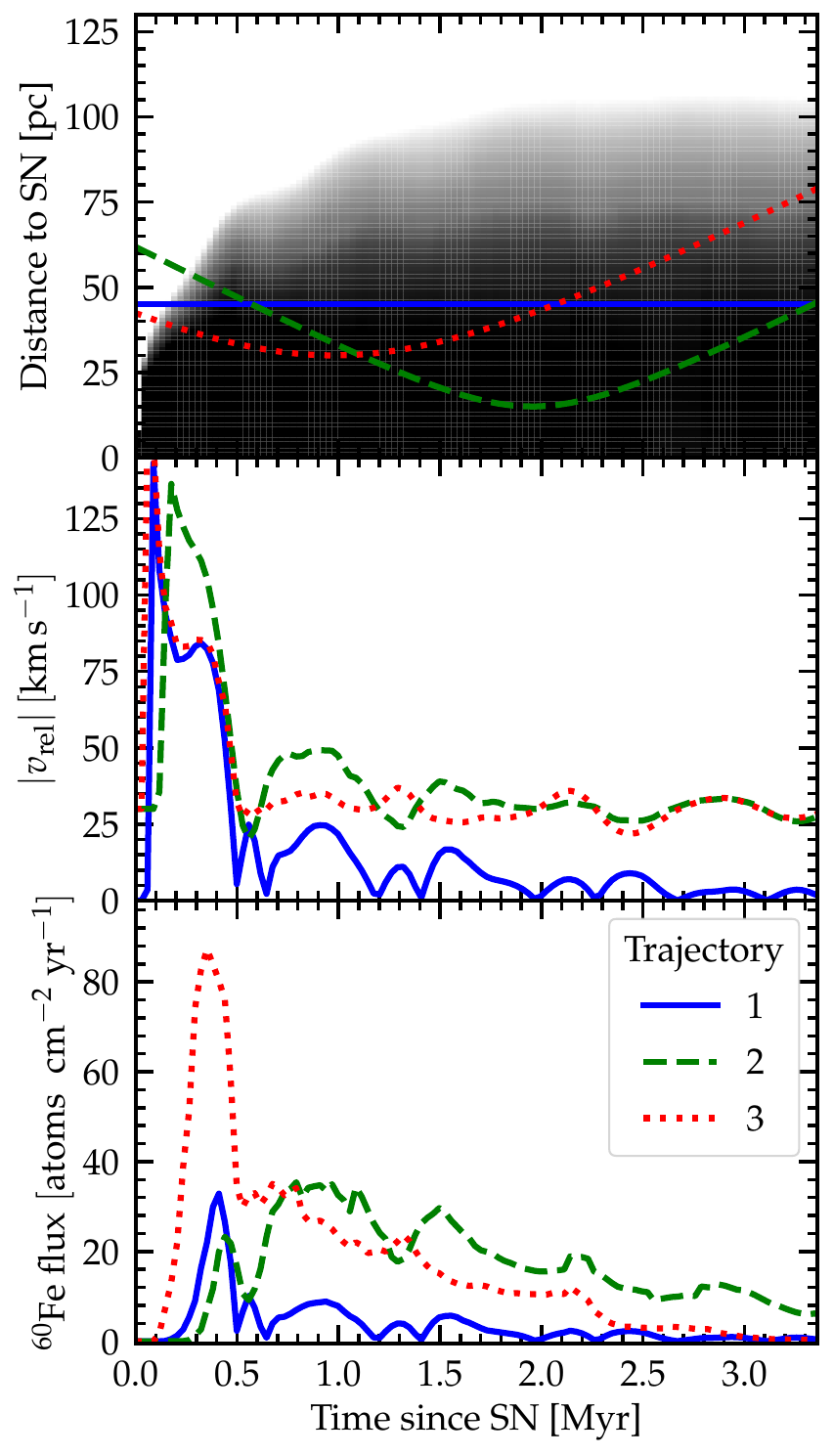}
    \caption{Three possible trajectories of the Solar System in the proximity of the SN and the resulting \Fe{} fluxes. The first trajectory (shown in blue) corresponds to a static observer positioned at 40 pc from the SN. The second (green) and the third (red) ones are linear trajectories piercing the ejecta with impact parameter of 15 and 30 pc, respectively. In these two cases, the Solar System is moving with velocity 30 \kms{} relative to the SN. The top panel is equivalent to the right panel of Figure \ref{fig:Fe60_flux_peak_ref_density} with three linear trajectories plotted in this phase space. The middle panel shows the absolute relative velocity between the Solar System and the ISM along these trajectories. The bottom panel shows the \Fe{} flux onto the Solar System assuming $M_{\rm ^{60}Fe,tot}  \, f_\mathrm{^{60}Fe} = 10^{-6}\, \rm M_\odot$ and including radioactive decay similarly to Figure \ref{fig:flux_well_mix_Fe60}. The trajectories are shown for the \textsf{high\_res\_n01} simulation.}
    \label{fig:trajectory_of_the_solar_system}
\end{figure}

As can be seen in the right panel of Figure \ref{fig:Fe60_flux_peak_ref_density}, our simulations predict that the bubble where the \Fe{} ejecta have a significant concentration expands only during the first $\sim 1$ Myr and then it remains almost static. Because the average radial velocity of the \Fe-rich gas after $\sim 1$ Myr becomes very low, the amplitude of the \Fe{} flux measured by a static observer will also be very low. This implies that neither in the case with numerical ejecta tracing nor in the well-mixed scenario can we produce \Fe{} pulses lasting longer than $\sim 1$ Myr. As we will show in the following, longer pulses can be constructed by considering a more general model where the observer is a moving observer rather than a static one. This model is motivated by the fact that the Solar System is moving relative to the local ISM. In order to properly estimate the amount of ejecta accreting onto the Solar System, we have to take into account the initial position and motion of the Solar System relative to the SN. There are no other parameters in our simplified spherically symmetric treatment of the ISM. The following analysis is based on our fiducial simulation, \textsf{high\_res\_n01}.

We consider three possible trajectories to display the variety of possible accretion histories one can expect in this geometry;
results appear in Figure \ref{fig:trajectory_of_the_solar_system}. 
\begin{itemize}
    \item The first one is a static position at 40 pc from the SN. 
In the bottom panel we see that the \Fe-flux history is markedly different from those in Figure \ref{fig:flux_well_mix_Fe60}, where ejecta were assumed to be well-mixed but observers were also static.  In the well-mixed case, the accretion histories are smooth, with a well-defined single peak.  Here, the limited and spherically asymmetric mixing of \Fe{} leads to a flux profile that shows a relatively sharp initial peak lasting $\sim 0.2$ Myr, but a long and irregular residual flux at a level $\sim 10-20$ per cent of the peak value over $\sim 3$ Myr.
This behaviour reflects the \Fe{} distribution pictured in the bottom panels of Figure \ref{fig:hydrogen_and_Fe60_profiles_3times}. The ejecta reach this radius rapidly and then stay concentrated in a bubble of radius $\sim100$ pc. Thus, the accretion in this scenario will have a rapid onset with a gradual falloff as the velocity decreases and the \Fe{} decays. This is shown in blue solid curves in Figure \ref{fig:trajectory_of_the_solar_system}.

\item The second and the third trajectories are linear inward trajectories that pierce through an almost static ejecta bubble with a velocity of 30 \kms{} and impact radius of 15 and 30 pc. Since we know that the Solar System currently moves relative to the local ISM at $\sim 26$ \kms{} \citep[e.g.][]{2012A&A...544A.135X}, we can safely consider the configurations with the Solar System moving with speed of a few $10$ \kms{} ($\sim10$ pc Myr$^{-1}$) when it crosses the ejecta. We see that the relative velocity can create yet more variety in the \Fe-flux profile, as proposed by \citet{Wallner2016}.  Trajectory 2 comes nearest to the explosion site and thus has a long duration in the \Fe{}, reflected in a somewhat irregular but sustained flux profile with a slow decline and a timespan of least $\sim 3$ Myr.  Trajectory 3 has a larger impact parameter relative to the explosion site, and the \Fe-flux history is between that of the trajectories 1 and 2: there is a pronounced peak, and then a long gradual decline that spans $\sim 40$ per cent down to $\sim 10$ per cent of the peak value over $\sim 3$ Myr.
\end{itemize}

These trajectories show that one can generate peaks of diverse size and intensity, and have implications for interpreting the \Fe{} data. It is noteworthy that in all three cases, there is only one peak: sharp and well-defined in trajectories 1 and 3, and broad in trajectory 2. Thus, within the context of our models, the two well-defined \Fe{} pulses seen in the \citet{Wallner742} data would require two SN events.  It remains for future work to investigate whether a single explosion into non-uniform media or a more complex trajectory could produce two peaks for some observers.

The extended flux after the sharp peaks in trajectories 1 and 3 may help account for the detection of \Fe{} in recent samples.  \citet{Koll2019} reports \Fe{} in modern Antarctic ice, while \citet{Wallner2020} sediment data show that a \Fe{} flux extends from the present back to $\sim 33$ kyr ago.  The flux levels are $\sim 10$ per cent of the peak at 3 Myr ago, broadly consistent with the dropoff seen in trajectory 1.

Finally, the broad trajectory 2 flux profile shows that long time-scales are possible if the observer is moving.  This offers a potential explanation for the $> 1 \ \rm Myr$ duration of the pulses seen in the sediment measurements of \citet{Ludwig2016} and \citet{Wallner2016}.

\subsection{Peak complexity and agreement with data}
\label{sec:data}

The results of our fiducial simulation, \textsf{high\_res\_n01}, presented in Figure \ref{fig:trajectory_of_the_solar_system} show that even in the scenario with a single SN in an homogeneous ISM the ejecta flux can have long  $\sim0.5-2$ Myr peaks with complex $\sim0.1$ Myr sub-structure. However, these short time-scale features are unlikely to be observable due to the limited time-resolution of the sediment data. The shapes of these peaks are very different compared with shorter and smoother peaks produced in the well-mixed scenario shown in Figure \ref{fig:flux_well_mix_Fe60}. This difference is caused by the generation of acoustic oscillations inside the SN bubble that are discussed in \S\ref{subsec:oscillations}. The commonality of the fluxes in Figure \ref{fig:trajectory_of_the_solar_system} along the three trajectories is that the onset of the accretion rate is rapid due to the headwind and typically higher concentration at earlier time, while the other side of the peaks is less steep due to the reversed conditions. This effect is further enhanced by the decay of the ejecta elements.

\begin{figure}
    \centering
    \includegraphics[width=0.97\linewidth]{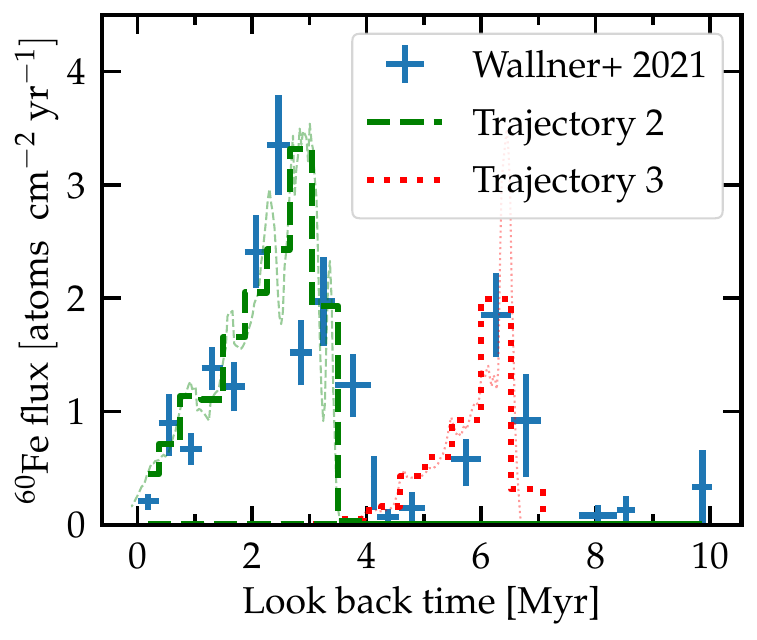}
    \caption{The incorporation rate of \Fe{} from \citet{Wallner742} with two separate trajectories shown in the top panel of Figure \ref{fig:trajectory_of_the_solar_system} (trajectories 2 and 3). The green dashed and red dotted trajectories are shifted in time by 3.8 and 6.8 Myr and are normalized assuming $M_{\rm ^{60}Fe,tot}  \, f_\mathrm{^{60}Fe}  = 1.0 \times10^{-7}  \, \rm M_\odot$ and $0.4\times 10^{-7}  \, \rm M_\odot$ (see \S\ref{sec:data} for details). Thin lines correspond to the time resolution of our simulation and repeat those in the bottom panel of Figure \ref{fig:trajectory_of_the_solar_system}, and thick lines mimic the time resolution of the data. Note that time axis is reversed compared to other figures. }
    \label{fig:wallner}
\end{figure}

There are multiple experiments measuring \Fe{} density in sediments that one can compare to. Here we only consider \citet{Wallner742} result that exhibits two peaks and show them in Figure \ref{fig:wallner}. We can reproduce these peaks separately by joining the second and the third linear trajectories described in \S\ref{sec:trajectories} and shifting them in time by 6.8 and 3.8 Myr accordingly, i.e. assuming two SNe separated in time by 3 Myr. The resulting evolution of \Fe{} flux from the second and third trajectories is shown in Figure \ref{fig:wallner} with the same time resolution as in the observed data.

The amplitude of the peaks are tuned based on the following assumptions. The conversion of accretion rate to observed signal in the sediments is regulated by the composition of three factors \citep{Fry2015}: \Fe{} mass in ejecta $M_{\rm ^{60}Fe,tot}$, which ranges from $10^{-6}$ to $10^{-3}\, \rm M_\odot$ for the Core-Collapse supernovae of different masses \citep{2006ApJ...647..483L}; the uptake in the Fe-Mn crust $U_\mathrm{^{60}Fe}$ whose estimates range from a per cent to a unity; and the fraction of \Fe{} entrained in dust $f_\mathrm{^{60}Fe, dust}$, which is order of $\sim 0.01$ \citep{Fry2015}. Because we do not make a clear distinction between how much \Fe{} is entrained in the blastwave in the form of dust and in the form of gas, we opted to combine the uptake $U_\mathrm{^{60}Fe}$ and dust fraction $f_\mathrm{^{60}Fe, dust}$ into a single parameter, $f_{\rm ^{60}Fe}$ (see equation \ref{eq: flux_Fe_well_mixed}), which characterises the overall \textit{efficiency} of \Fe{} deposition onto the Earth surface. For the trajectories 2 and 3 shown in Figure \ref{fig:wallner} we use $M_{\rm ^{60}Fe,tot} f_\mathrm{^{60}Fe}  = 1.0\times 10^{-7} \, \rm M_\odot$ and $0.4 \times 10^{-7} \, \rm M_\odot$, respectively.

In reality, one does not expect  the Solar System's trajectory to be linear for $\sim10$ Myr and we know its velocity and acceleration in the galactic disk \citep[e.g.][]{2012A&A...544A.135X}. Also, the ISM morphology is an actively developing field \citep[e.g.][]{2015ApJ...815...67K} and is also known to be more sophisticated than what we assumed in this study. Thus, our study has a potential to be expanded by adopting more realistic trajectory and ISM configurations.

\section{Discussion}

\subsection{Major differences with the well-mixed scenario and ram pressure}
There are two distinct features that are present in the scenario with \Fe{} well-mixed within the shock, and that are absent in the simulations with explicit ejecta tracing. 

Firstly, if \Fe{} is well-mixed it travels in the ISM as far as the blastwave, meaning that the Solar System can be located further than $100$ pc from the SN but still encounter the ejecta. In the full simulation, the distribution of \Fe{} is concentrated next to the SN and does not go beyond $\sim100$ pc.

Secondly, in the well-mixed scenario the peak in \Fe{} density coincides with the shock and consequently with the peak in ram pressure. The reduction in the size of the heliosphere reflects an equilibrium between the solar wind and supernova blast ram pressures \citep{Fields2008}. Thus, one can argue that ram pressure temporarily reduces the size of the heliosphere and assists the deposition of \Fe{} onto the inner parts of the Solar System. After its sudden onset, the blast ram pressure decreases with time, as can be inferred from the middle panel of Figure ~\ref{fig:Fe60_flux_peak_ref_density}. This will lead to a corresponding retreat of the heliosphere over the same time-scale.

This picture does not hold in our simulation with \Fe{} tracing because the peak in \Fe{} accretion is not synchronised with the peak in ram pressure. Moreover, since our selected trajectories are relatively far from the supernova, the peak ram pressure when the shock passes through reaches few $10^{-11}\;\mathrm{dyne\,cm^{-2}}$, which is orders of magnitude lower than the solar-wind pressure at 1 au which is  $\sim2\times10^{-8}\;\mathrm{dyne\,cm^{-2}}$. Thus, the necessity to transport the ejecta into the Solar System by dust remains. In any case, the observed $>1$ Myr duration of \Fe{} peak in the Earth sediments suggests that the ram pressure alone should not play a controlling role in the accretion rate since its peak is shorter in any of the scenarios.

\subsection{Comparison with previous work}

Our work builds on several previous studies. Our findings are generally in good agreement on points of overlap, but our work is complementary, being different in scope and focus. \citet{Fry2020} study the gas-phase \Fe{} ejecta in a single remnant, using `1.5D' models that approximate the effects of turbulent mixing.  They also find that the ejecta lag considerably behind the forward shock, even in the presence of mixing.

Two groups performed 3D hydrodynamics simulations of supernova ejecta dispersal, both on larger scales than ours in both space and time, and both groups simulated many supernovae together. \citet{Breitschwerdt2016} and \citet{Schulreich2017} simulated the formation of the Local Bubble. They used an adaptive mesh refinement code to model a 3-kpc cube, with a highest resolution of $0.7$ pc. They modelled 19 supernova explosions and included passive tracers to follow \Fe\ and the resulting entrained signal. They considered observers at rest, and found that the ejecta propagation for each individual supernova generally leads to rather narrow signals $< 1$ Myr, which is in line with the results in this work for the case of a static observer. Longer pulses could arise if the blasts from two explosions are close enough in time to overlap.

\citet{Fujimoto2020} focussed on Galactic scales, modelling a Milky-Way-like spiral galaxy, following the disk evolution in an adaptive mesh refinement code with N-body dark matter. They included \Fe\ tracers created by supernova sources. They found that spiral density waves created kpc-scale bubbles enriched in SN ejecta. They looked at trajectories of stars within these large bubbles, and present example \Fe\ flux histories for sample observer on stars chosen to have Sun-like orbits. Their focus was on the intensity of the  \Fe\ flux, and found that a small but non-negligible fraction of stars ($\sim 10-20$ per cent) could encounter fluxes as large or larger than that seen in the deep ocean.

Our work differs from these important studies in that we focus on smaller scales and investigate the effects of individual supernovae. Our treatment of the observer's velocity is similar in some ways to \citet{Fujimoto2020}, but here we focus not only on the observed \Fe-signal intensity but also the duration, and we make detailed comparison with observed pulses. We also consider effects of diffusion not examined in earlier work.

Furthermore, the simulations in \citet{Breitschwerdt2016} and \citet{Fujimoto2020} were run using Eulerian grid codes including adaptive mesh refinement, whereas we employed a Lagrangian SPH code. Unlike  grid-based fluid solvers, in SPH each gas particle represents a fluid element so that tracing \Fe{} ejecta becomes trivial. The spatial resolution in SPH is proportional to particles' smoothing kernels; higher (lower) density regions are made up of a greater (lower) numbers of SPH particles that have smaller (larger) kernels resulting in naturally adaptive resolution \citep{Price2012}. As far as the accuracy in the ST blastwave test, it is possible to converge towards the known analytical solution in both SPH and grid-based codes \citep{Tasker2008,price2018,2020arXiv201203974B}.

\subsection{Caveats and limitations} 

Compared to the real ISM, the initial conditions in our simulations are subject to several assumptions, the most significant of which are the lack of density inhomogeneities and gas turbulent motions. The impact of this more complex physics has been extensively studied in the past. \cite{Martizzi2015} and \cite{Haid2016} investigated the SN evolution in a medium with the lognormal density distribution expected from supersonic turbulence. They showed that the blastwave's asymptotic radial momentum is largely independent of the turbulent structure, while the blastwave (effective) radius can become larger because the SN bubble escapes through low-density channels. Similar results were found by \cite{Ohlin2019} who ran high-resolution simulations of SNe in turbulent environments in which the turbulence was generated using a randomised forcing field with a given spectrum in Fourier space. Based on the aforementioned studies, we expect that our main findings will be valid also in the presence of turbulence and density inhomogeneities. However, the mixing between \Fe-rich and \Fe-free gas around the contact discontinuity will be enhanced and \Fe-signal profiles will acquire more complex shapes.  

Another caveat of our simulations is the lack of ISM magnetic fields. While magnetic fields generally have a mild impact on the properties and evolution of the SN blast \citep[e.g.][]{2015ApJ...815...67K, Iffrig2015}, their influence on dust particles -- the most likely carrier of \Fe{} ejecta to Earth -- is far more dramatic. \cite{Fry2020} traced trajectories of individual dust grains in a magnetized ISM. They included the effects of dust sputtering, drag, and charging.  As in prior work, they showed that dust grains decouple from the gas, and they further showed that the presence of magnetic fields in the shocked ISM restricts the movement of the grains by either reflecting or trapping them within the SN blast. The propagation of \Fe{}-bearing dust grains to Earth is thus distinct from the spread of the gas-phase ejecta, and the long stopping time-scale may explain the long \Fe{} time-scale. 

In order to be able to carry \Fe{} to Earth, supernova dust grains first need to survive.  This larger question has sparked intense study, with the theory and observations suggestion sputtering and shattering points to substantial grain destruction \citep[e.g.][]{Bianchi2007,Micelotta2016}, while significant grain production and survival is suggested by, e.g., the observations of dusty galaxies at high redshifts \citep{Bertoldi2003,Maiolino2004}, and the presence of supernova-produced pre-solar grains in meteorites \citep[e.g.][]{Zinner1998,Gyngard2018}. All in all, the issues of dust survival and trajectories in a magnetized ISM require more investigation, ideally with a full 3D hydrodynamical simulation including magnetic fields and proper treatment of dust, which is beyond the scope of this work. 

\section{Conclusions}

We used 3D hydrodynamical simulations of isolated supernovae in the ISM of uniform initial density to study the propagation of \Fe{} entrained in the gas. The tracers of \Fe{} are assumed to be either in the gas phase, or in the form of dust which is at rest with respect to the fluid. We considered two models for mixing of \Fe{}. For the first model, in which the \Fe{} ejecta are assumed to be well-mixed in the shocked ISM, our main findings are as follows:

\begin{itemize}

\item The observed \Fe{} signals can have widths from $\sim 0.1$ to $\sim 1$ Myr depending on the density of the ISM and how far from the SN the observer is located. The range of distances that produce a realistic signal is roughly from $10$ to $200$ pc, which is in agreement with previous studies. 
 
\item The inclusion of gas radiative cooling has several effects on the observed \Fe{} signal. First, compared to the predictions based purely on the ST analytical solution, following the time of shell formation the shock travels slower. This reduction in speed delays the onset of \Fe{} accretion, which  generally leads a decrease in the amplitude of the \Fe{} accretion rate. Second, the flux shapes become broader because at a lower relative speed it takes the shock more time to pass through observer. None the less, it remains very difficult to produce realistic signals with widths larger than $1$ Myr.

\end{itemize}

In the second model, the ejecta are numerically traced with a passive scalar field. Our main findings are:

\begin{itemize}

\item  The most distinct feature of this model compared to the well-mixed one is that \Fe{} significantly lags behind the shock. Consequently, the moment when the supernova shock passes by the Solar System and the ram pressure on the heliosphere reaches its maximum significantly precedes the moment when the surrounding gas becomes enriched with the entrained ejecta and \Fe{} in particular. This questions the mechanism of injecting \Fe{} into the inner Solar System (as discussed in \citealt{Fields2008}). It also favours the scenario in which \Fe{} is delivered onto Earth by dust grains, since they can overcome the solar-wind pressure. We note, however, that we considered only a singular supernova, while a combination of two consequent supernovae can create the condition for simultaneous high ejecta concentration in the surrounding gas and high ram pressure. 

\item A lag between the arrival of the forward shock and the \Fe{} may also have implications for the timing of possible biological damage triggered by different aspects of the explosion and blast \citet{Thomas2016,Melott2017,Melott2019,Melott2011}.  The effects of cosmic rays would commence with the arrival of the forward shock, and thus would precede the \Fe{} arrival.  If SN(e) were responsible for earlier extinction events such as the end-Devonian and deposited observable radioisotopes \citep{Fields2020}, such a lag would be present there too.

\item We investigated the impact of the Solar trajectory in the ISM and demonstrated that different trajectories lead to very different accretion histories of entrained material onto the heliosphere. In particular, we showed that even a single supernova explosion in the homogeneous ISM in combination with a favourable trajectory can create a few-Myrs impulse that is consistent with recent observations \citep{Ludwig2016,Wallner2016,Wallner742}. This result provides additional support to the theory that the \Fe{} signal was originated from near-Earth SNe. With two supernovae we can mimic the observed double-peak structure from \citet{Wallner742} as shown in Figure \ref{fig:wallner}.

\item For two of the trajectories in Figure \ref{fig:trajectory_of_the_solar_system}, \Fe{} shows a sharp peak followed by a extended residual flux over $\sim 3$ Myr at levels ranging from $10-40$ per cent of the peak value,  This is suggestive of the reduced but non-zero \Fe{} flux reported in modern Antarctic snow \citep{Koll2019} and extending back to $\sim 33$ kyr ago in marine sediments \citep{Wallner2020}.

\item Irrespective of assumed trajectory, all accretion rate histories generated in this model have a similar general shape with a rapid onset at first and then a gradual decrease, as seen in Figure \ref{fig:trajectory_of_the_solar_system}. Its is remarkable that the observed peaks shown in Figure \ref{fig:wallner} somewhat resemble this shape as well.

\item The single-peaked nature of the \Fe{} histories implies that the two distinct peaks seen in \citet{Wallner742} indeed require two supernovae.  Whether single explosions in non-uniform ISM models could reproduce such features remains a question for future work.

\end{itemize}

The future development of the numerical methods presented in this paper may include more sophisticated initial conditions for the inhomogeneous ISM, multiple supernova explosions, more realistic trajectories of the Solar System within the local Galactic potential. Ultimately such a hydrodynamic simulation should be run in conjunction with the simulation of the decoupled dust propagation in order to fully and self-consistently track the distribution of the ejecta in the galactic disk.  These models also provide
the initial conditions for studies of the propagation of supernova plasma and dust into the Solar System.

In closing, we note that \citet{Wallner742}
has reported the detection of \Pu{244}
in a ferromanganese crust,
with less time resolution but in layers
reaching to $\sim 10$ Myr.
This exciting result probes
the origin of the heaviest elements
in the {\em r}-process.
\citet{Wang2021} and Wang et al~(in preparation) argue that
the measured \Fe/\Pu{244} ratio requires
that both recent supernovae were unusual,
or that a kilonova event $> 10$ Myr ago
seeded the (proto)-Local Bubble with
\Pu{244} and other {\em r}-process
species.
These scenarios cry out for further investigation that builds on the present work.

\section*{Acknowledgements}
We thank Josh Borrow, Matthieu Schaller, and Joop Schaye for their useful comments.
BDF is grateful for many discussions
with longtime collaborators whose
insights have informed this work,
especially John Ellis, Adrienne Ertel,
Brian Fry, and Jesse Miller.
EC is supported by the funding from the European Union's Horizon 2020 research and innovation programme under the Marie Skłodowska-Curie grant agreement No 860744.
AAK is supported by William D. Loughlin fellowship. This work used the DiRAC@Durham facility managed by the Institute for Computational Cosmology on behalf of the STFC DiRAC HPC Facility (www.dirac.ac.uk). The equipment was funded by BEIS capital funding via STFC capital grants ST/K00042X/1, ST/P002293/1, ST/R002371/1 and ST/S002502/1, Durham University and STFC operations grant ST/R000832/1. DiRAC is part of the National e-Infrastructure. This work is partly funded by Vici grant 639.043.409 from the Dutch Research Council (NWO).  B.D.F. was supported in part by the NSF under grant number AST-2108589. CC acknowledges the support by the Dutch Research Council (NWO Veni 192.020). The research in this paper made use of the \textsc{swift} open-source simulation code (\url{http://www.swiftsim.com}, \citealt{2018ascl.soft05020S}) version 0.9.0; all data from the simulations were processed with the help of \textsc{swiftsimio} \citep{Borrow2020swiftsimio}.

\section*{Data Availability}

The data underlying this article will be shared on reasonable request to the corresponding author.



\bibliographystyle{mnras}
\bibliography{paper} 




\appendix

\section{Resolution tests}
\label{ap:resolution_tests}
\begin{figure}
    \centering
    \includegraphics[width=0.99\linewidth]{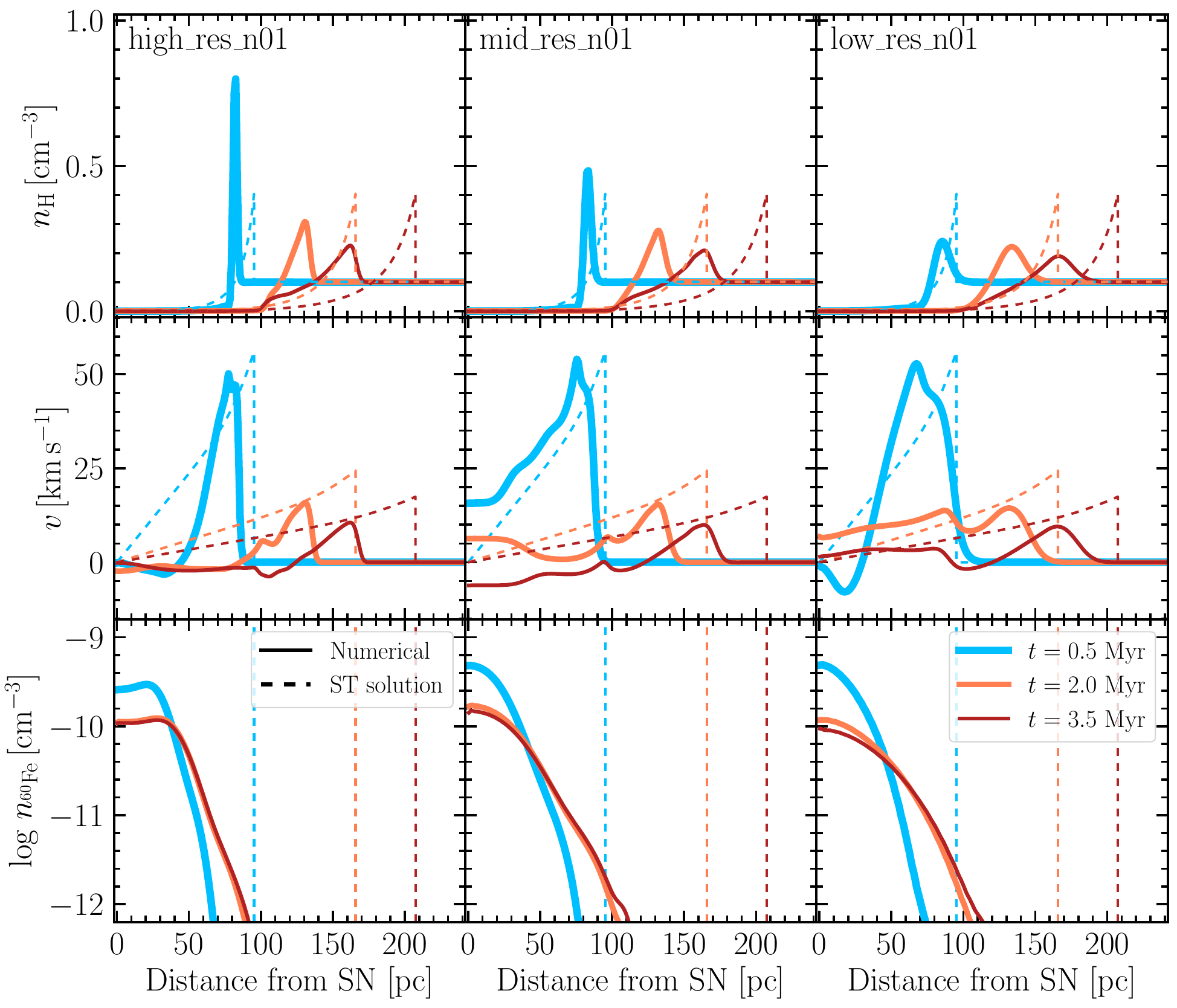}
    \caption{Convergence with gas-particle mass. The resolution increases from right to left. Shown are the average radial profiles of hydrogen number density (\textit{top}), gas radial velocity (\textit{middle}), and number density of \Fe{} (\textit{bottom}) in the simulations with gas-particle mass $M_{\rm gas} = 3 \times 10^{-2} \, \rm M_\odot$ (\textsf{high\_res\_n01}, \textit{left}), $M_{\rm gas} = 24  \times 10^{-2} \, \rm M_\odot$ (\textsf{middle\_res\_n01}, \textit{middle}), and $M_{\rm gas} = 192 \times 10^{-2}  \, \rm M_\odot$ (\textsf{low\_res\_n01}, \textit{right}). The profiles are displayed at three consecutive times: $0.5$ Myr (light-blue), $2.0$ Myr (orange), and $3.5$ (dark-red) Myr. The mean ISM density is $n_{\rm H} = 0.1$ cm$^{-3}$ in all cases. The width of radial bins is $1$ pc. For reference, we use the thin dashed curves to indicate the ideal density profiles (\textit{top}), velocity profiles (\textit{middle}) and forward shock positions (\textit{bottom}) in the ST solution. To compute the ejecta density we assumed $M_{\rm ^{60}Fe, tot}=10^{-4} \, \rm M_\odot$ and no radioactive decay.}
    \label{fig:convergence}
\end{figure}

In Figure \ref{fig:convergence} we show how the gas density, gas radial velocity, and \Fe-ejecta density depend on the resolution in our simulations with mean ISM density $n_{\rm H} = 0.1$ cm$^{-3}$. We display average radial profiles of these three fields at three consecutive times: $0.5$, $2.0$, and $3.5$ Myr. All radial bins have the same width of $1$ pc and the averages are computed as in equation (\ref{eq: n_Fe60_average}). The run at our fiducial resolution ($M_{\rm gas} = 3 \times 10^{-2} \, \rm M_\odot$) is shown in the left column. The SPH particle mass increases by a factor of $8$ between the left and middle column, and by another factor of $8$ between the middle and right column. For each factor-of-eight increase in SPH mass, the SPH smoothing length increases by a factor of $2$. The latter can be seen in the width of the shock, which becomes larger from left to right. However, the shock position is unaffected by the change in the resolution. Importantly, the radial profiles of \Fe{} look very similar too, indicating that the temporal and spatial evolution of \Fe{} ejecta is not sensitive to changes in the resolution.

\section{Uncertainties in distribution of ejecta}

\subsection{Effect of diffusion constant}
\label{appendix: effect_of_diffusion}

\begin{figure}
    \centering
    \includegraphics[width=0.99\linewidth]{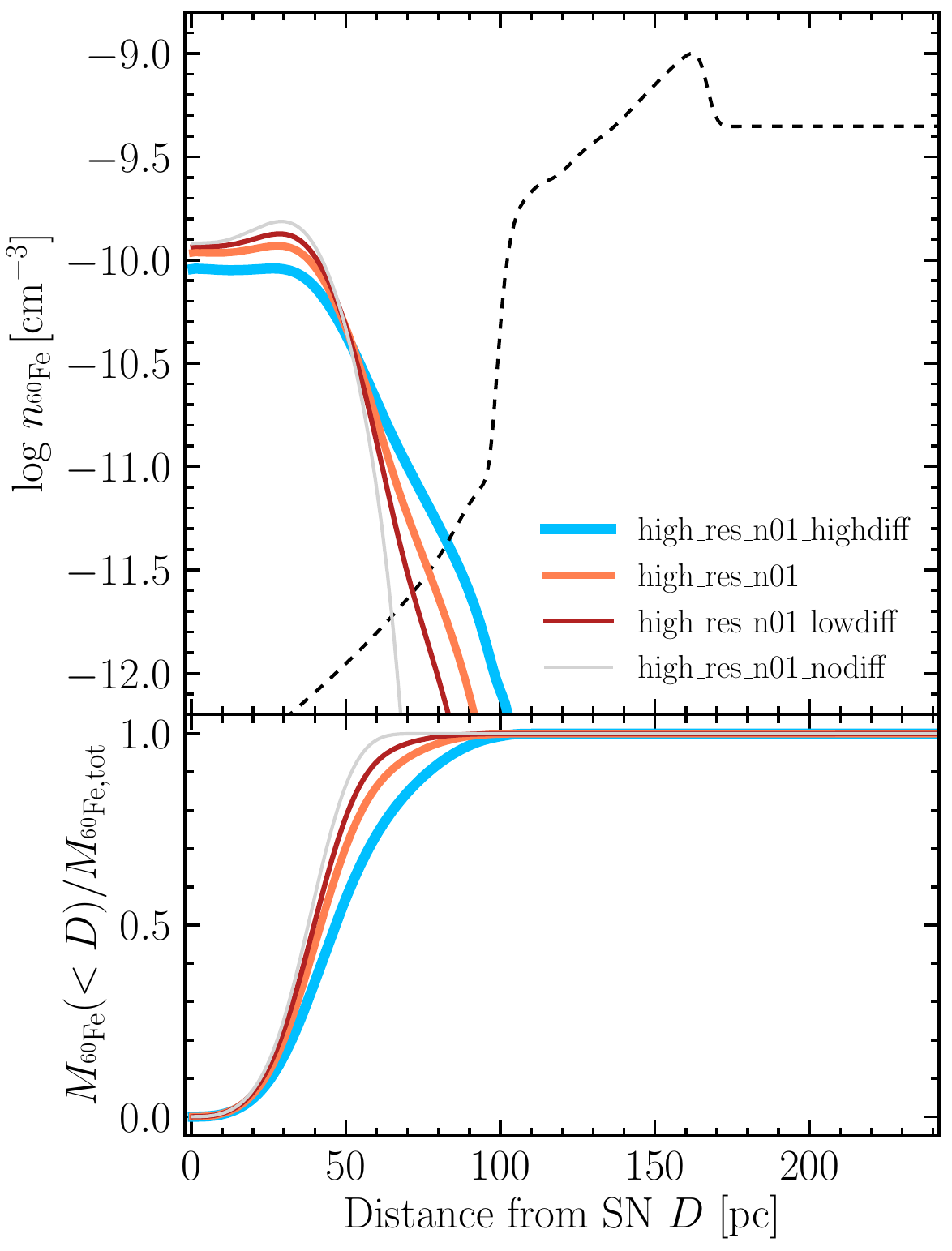}
    \caption{\textit{Top:} Radial profiles of \Fe{} density shown at $t=3.5$ Myr in simulations with different values of the diffusion constant $C_{\rm D}$: $C_{\rm D}=0.20$ (\textsf{high\_res\_n01\_highdiff}, light-blue), $C_{\rm D}=0.10$ (\textsf{high\_res\_n01}, orange), $C_{\rm D}=0.05$ (\textsf{high\_res\_n01\_lowdiff}, dark-red), and $C_{\rm D}=0.00$ (\textsf{high\_res\_n01\_nodiff}, light-grey). For reference, we also show the radial profile of gas mass density from \textsf{high\_res\_n01} at $t=3.5$ Myr, which is displayed as a black dashed curve. The mass-density profile is the same in all the four simulations. The units of the mass density are chosen such that the curve reaches its maximum at a value of $10^{-9}$ in the Y axis. \textit{Bottom:} The cumulative fraction of \Fe{} mass as a function of distance from the SN, for the same four runs as in the above panel. The mean ISM density is $n_{\rm H} = 0.1$ cm$^{-3}$ in all cases.  The size of all radial bins is $1$ pc. To compute the ejecta density we assumed $M_{\rm ^{60}Fe, tot}=10^{-4} \, \rm M_\odot$ and no radioactive decay.}
    \label{fig:diffusion}
\end{figure}

In Figure \ref{fig:diffusion} we illustrate the effect of varying the diffusion constant $C_{\rm D}$, which is one of the free parameters in our simulations. The top panel shows radial density profiles of \Fe{} ejecta at time $t=3.5$ Myr counted since the SN went off. We consider four simulations, all at our fiducial resolution and mean ISM density $n_{\rm H}=0.1$ cm$^{-3}$, whose diffusion constants are equal to $0.00$ (no diffusion), $0.05$ (low diffusion), $0.10$ (fiducial case), and $0.20$ (high diffusion). For reference, we also show the radial profile of gas mass density at $t=3.5$ Myr, which is the same in these four simulations. Expectedly, we find that increasing (decreasing) the diffusion constant pushes the \Fe{} ejecta to larger (smaller) radii. 

The bottom panel displays the cumulative fraction of \Fe{} mass, $M_{\rm ^{60}Fe}(<D)/M_{\rm ^{60}Fe, tot}$, shown as a function of distance from the SN, $D$. The mass $M_{\rm ^{60}Fe}(<D)$ is the \Fe{} mass within a sphere of radius $D$ whose origin coincides with the position of the SN. The total \Fe{} yield, $M_{\rm ^{60}Fe, tot}$, is equal to $10^{-4} \, \rm M_\odot$. We find that regardless of what the value of $C_{\rm D}$ is, $90$ per cent of the \Fe-ejecta mass is always concentrated within a sphere of $\approx 80$ pc, and nearly $100$ per cent is reached at a distance of $\approx 100$ pc. These values are significantly smaller than the position of the blastwave at this time, $\approx 160$ pc. At a fixed distance, changing the diffusion constant by a factor of $2$ results in variations in the distribution of the mass fraction of \Fe{} by up to $\approx 20$ per cent with the overall shape of the distribution remaining the same. We hence conclude that our results are robust and only weakly dependent on the exact value of the diffusion constant $C_{\rm D}$.

\subsection{Impact of initial distribution of ejecta} 
\label{appendix: effect_of_ICs}

We investigate the impact of alternative initial ejecta distributions. In the fiducial run, we assumed that the \Fe{} ejecta are distributed uniformly in a sphere of 5 pc and initially are carried by 57 SPH particles, which is the closest integer to the expected number of neighbours in a particle's SPH kernel, $N_{\rm ngb} = 57.28$ The fact that the number of particles (initially) enriched with \Fe{} is rather small should not play a big role in that, owing to our subgrid diffusion model, after a few time-steps this number will increase exponentially. We verify that that is indeed the case by considering a run variation where the \Fe{} ejecta and SN energy are distributed uniformly in a smaller sphere, of size 2.5 pc comprising just 8 particles. Another interesting test we consider is to assume a different scaling of \Fe{} density with distance from the SN but leave the distribution of SN energy unaffected. Since our simulations predict that the concentration of \Fe{} peaks close to the centre of the SN bubble, it makes sense to use an initial \Fe-density distribution with similar spatial behaviour. We create such a variation by distributing the ejecta among the 57 particles inside the 5-pc sphere with a density profile of $1/r^2$ where $r$ is the radial distance (the energy has the same, uniform distribution as in the fiducial run). \\

\begin{figure}
    \centering
    \includegraphics[width=0.99\linewidth]{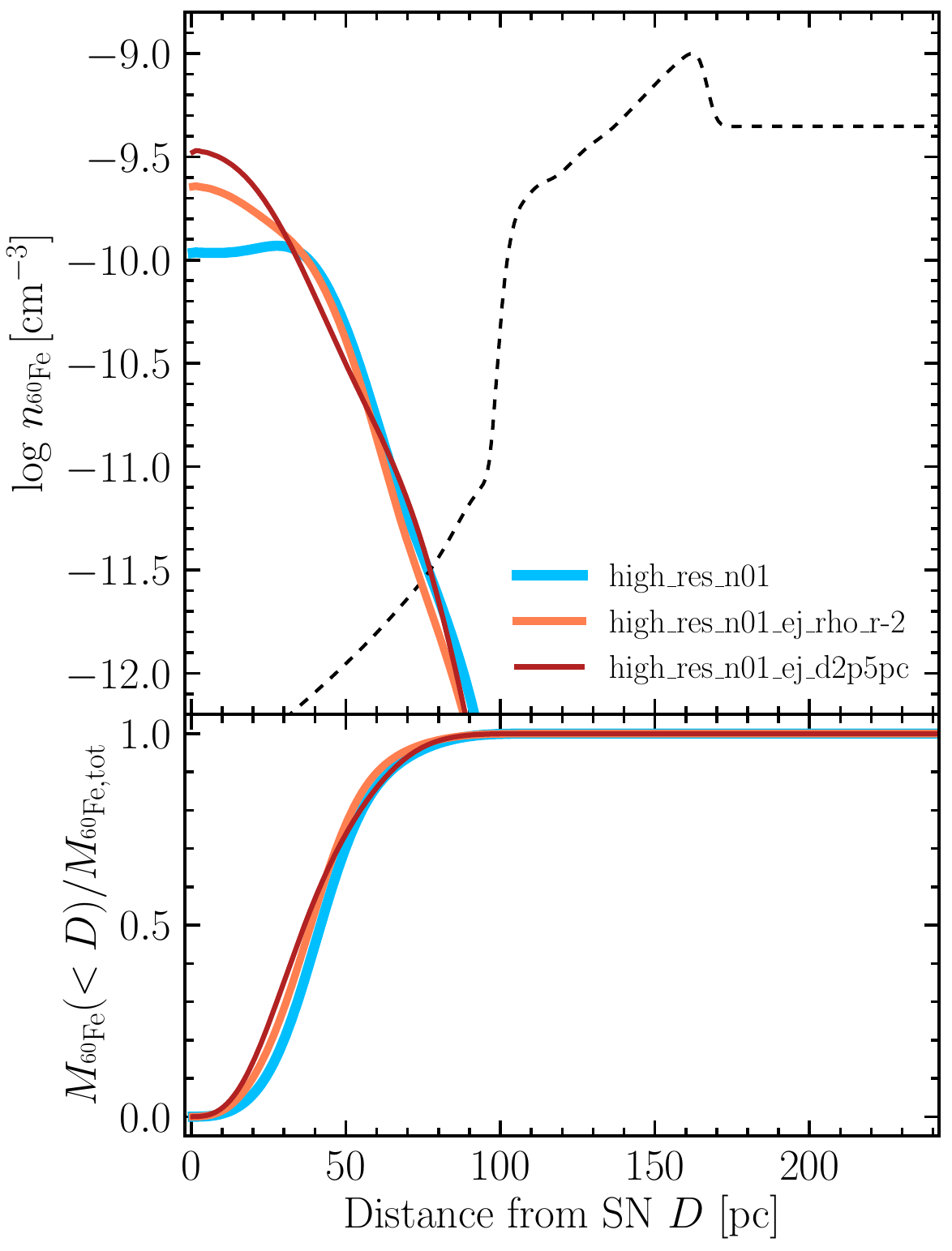}
    \caption{Same as Figure \ref{fig:diffusion}, but showing the effect of variations in the initial ejecta distribution. The fiducial run (\textsf{high\_res\_n01}), with the \Fe{} and SN energy distributed uniformly in a 5-pc sphere, is shown in light-blue. The orange curve  (\textsf{high\_res\_n01\_ej\_rho\_r-2}) shows the run where the \Fe{} ejecta are distributed in a 5-pc sphere with a $1/r^2$ density profile (the energy is still distributed uniformly). The dark-red curve (\textsf{high\_res\_n01\_ej\_d2p5pc})  shows the run where the \Fe{} and SN energy are uniformly distributed in a 2.5-pc sphere.}
    \label{fig:diff_initial_dist_of_Fe60}
\end{figure}

In Figure \ref{fig:diff_initial_dist_of_Fe60} we show radial profiles of ejecta density (top panel) and the ejecta mass fraction contained within a sphere of a given radius (bottom panel), for the fiducial run (light-blue) and the two runs with the variations in initial ejecta distributions described above (displayed in orange and dark-red). The results are shown at time $t = 3.5$ Myr. Unsurprisingly, we find that the ejecta-density profiles look nearly fully converged at distances $D \gtrsim 25$ pc confirming that the initial number of particles enriched with \Fe{} and the initial \Fe-density profile have little impact on our final results. The variations we see in the innermost region of the SN bubble ($D \lesssim 10$ pc), which are within a factor of $\sim 3$ with respect to the reference run, are much smaller than the uncertainties in \Fe{} yields and the fraction of \Fe{} that is able to penetrate the Solar System (order(s) of magnitude). The reason we see them is because the lowest-density regions are resolved in SPH with just a handful of SPH particles, meaning that the \Fe{} density becomes very sensitive to the exact values of \Fe{} masses carried by those few particles.


\bsp	
\label{lastpage}
\end{document}